\newcommand{\be}{\begin{equation}}
\newcommand{\ee}{\end{equation}}
\newcommand{\ba}{\begin{eqnarray}}
\newcommand{\ea}{\end{eqnarray}}
\newcommand{\baa}{\begin{eqnarray*}}
\newcommand{\eaa}{\end{eqnarray*}}
\newcommand{\bb}{\begin{thebibliography}}
\newcommand{\eb}{\end{thebibliography}}

\newcommand{\bit}{\begin{itemize}}
\newcommand{\eit}{\end{itemize}}

\documentstyle[sprocl,epsf]{article}

\def\be{\begin{equation}}
\def\ee{\end{equation}}
\def\bea{\begin{eqnarray}}
\def\eea{\end{eqnarray}}

\begin{document}

\title{INTRODUCTION TO QCD SUM RULE APPROACH}

\author{A.V. RADYUSHKIN} 

\address{Physics Department, Old Dominion University,
Norfolk, VA 23529 \\ and \\
Theory Group, Jefferson Lab, Newport News, VA 23606} 

\maketitle\abstracts{In these lectures, I describe the  
techniques  used within  the QCD sum rule approach. 
The basic concepts of the approach are introduced using 
 a simple model of quantum-mechanical
oscillator in 2+1 dimensions. Then I discuss 
their field-theoretical extension and the construction
of the operator product expansion  for current correlators
in QCD. The calculation of static parameters is illustrated
on the example of 
sum rules in the vector and axial meson channels.
Finally,  the QCD sum rule 
calculation of the pion electromagnetic form factor
 is presented as an example of application of the method  
 to dynamic characteristics of the hadrons. }

\section{Introduction}

\subsection{Introductory remarks}\label{subsec:intremks}

QCD sum rules, invented more than 20 years ago by Shifman, Vainshtein
and Zakharov\cite{SVZ} are still a rather popular approach for theoretical
 study
of  the hadronic structure. Applied originally to the simplest
static hadronic characteristics, like masses, leptonic widths etc.,
they were also used to calculate much more complicated 
 things like hadronic wave functions and  form factors.

To begin with, I will briefly discuss the basic questions \\
$\bullet$ Why we need QCD sum rules? \\
$\bullet$ What are QCD sum rules? \\
$\bullet$ How the QCD sum rules are working?\\ 
using for illustration the QCD calculation of the pion form factor.
In subsequent sections, I will discuss in more detail
the machinery of QCD sum rules in several cases:
calculation of the properties of the lowest  resonances
in $\rho$-meson and $\pi/A_1$  channels,
and analysis of the pion form factor in 
various kinematical situations.

\subsection{ Pion form factor in QCD} 

\begin{figure}[t]
\mbox{
   \epsfxsize=12cm
 \epsfysize=4.5cm
  \epsffile{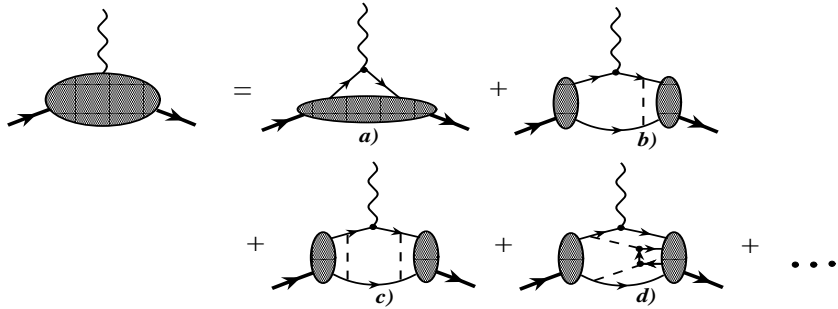}  }
{\caption{\label{fig:fact} 
 Structure of factorization for the pion form factor in QCD}}
\end{figure}

According to 
the factorization theorem\cite{er,lb}, 
for sufficiently large momentum transfers, one can represent the pion 
 form factor as a sum of 
 terms of increasing complexity (see Fig.1).
The first term (purely {\em soft}  contribution, Fig.\ref{fig:fact}a
 contains no 
short-distance (SD) subprocesses. The second term Fig.\ref{fig:fact}b contains a 
{\em hard} gluon exchange. There are also corrections to the hard term: 
higher order corrections Fig.\ref{fig:fact}c
  containing extra $\alpha_s$ factors 
and higher twist 
corrections Fig.\ref{fig:fact}d. In perturbative QCD one can calculate only 
the hard terms of this expansion.

The first result of perturbative QCD 
for exclusive processes is the prediction for the asymptotic behavior
of the pion electromagnetic form factor \cite{czs,er,farjack,bl}:
\be
F_\pi(Q^2)=\int_0^1 dx\int_0^1 \ t(xP,(1-x)P ; yP^\prime, (1-y)P^\prime, q) \ 
\varphi(x)\ \varphi(y) \ dy  \ ,  \label{eq:qcdff}
\end{equation}
where $\varphi(x)$ is the pion wave function giving the probability
amplitude to find the pion in a state where quarks carry
fractions $xP$ and $(1-x)P$ of its longitudinal 
momentum $P;$\, $q=P^\prime-P$ is the
momentum transfer to the pion
and $T(\ldots)$ is the perturbatively calculable 
short-distance amplitude.
The wave function $\varphi(x)$ accumulates long-distance information about
the pion structure and cannot be calculated within the pQCD approach.
Normally, some phenomenological assumptions are used 
about the form of this function depending on a single parameter $x$.
Still, there is a challenge to calculate it from first principles of QCD.

To calculate the soft contribution to the form factor 
one should know something
like the light-cone wave function $\psi(x,k_{\perp})$ depending also
on the transverse momentum $k_{\perp}$. I say ``something like'', 
because in principle it is impossible to prove that the soft term
reduces in QCD to the convolution  
\be
F(q^2) \sim \ \int \psi_{P}(x,k_{\perp}) \psi_{P}^*(x,k_{\perp}+xq)
\ d^2 k_{\perp} dx .
\end{equation}
In fact, one should treat the soft term as a nonfactorizable amplitude 
and the problem is to calculate it in some reliable way.
 
The QCD sum rules are just giving the method to calculate the 
nonperturbative hadronic characteristics incorporating the asymptotic
freedom property of QCD in a way similar to pQCD.

\subsection{Basic ideas of the QCD sum rule approach}

Among the existing approaches to the analysis of the
nonperturbative effects in QCD the most close to perturbative QCD
is the QCD sum rule method \cite{SVZ}. Let us formulate its basic ideas
within the context of the pion form factor problem.

It is evident that one cannot
directly study the soft contribution with the on-shell pions,
because then only long distances are involved. But perturbative
QCD can be applied in a situation when all relevant momenta
$q,$ $p_1,$ $p_2$ are spacelike and sufficiently large:
$|q^2|,$ $|p_1^2|,$ $|p_2^2|>1\,${\rm GeV}$^2.$
To describe the virtual pions one should use some
interpolating field, the usual 
(for the QCD sum rule practitioners) choice 
being the axial current $j_5^\alpha=
\bar d\gamma_5\gamma^\alpha u$. Its projection onto the pion state
$|P,\pi\rangle$ is proportional to
$f_\pi:$ 
\be
\langle0|j_5^\alpha|P,\pi\rangle=if_\pi P^\alpha.  \label{eq:axcur}
\end{equation}
{\em Via}\, the dispersion relation
\be
T(p_1^2,p_2^2,q^2)=\frac1{\pi^2}\int_0^\infty ds_1\int_0^\infty ds_2
 \ \frac{\rho(s_1,s_2,q^2)}{(s_1-p_1^2)(s_2-p_2^2)} \label{eq:dr}
\end{equation}
one can relate the amplitude $T(p_1^2,p_2^2,q^2)$ to its time-like counterpart
$\rho(s_1,s_2,q^2)$ containing the double pole term
\be
\rho_\pi(s_1,s_2,q^2)=\pi^2 f_\pi^2 \delta(s_1-m_\pi^2)\delta(s_2-m_\pi^2)
		      F_\pi(Q^2)              \label{eq:ropi}
\end{equation}
corresponding to the pion form factor. However, the axial current
has nonzero projections onto other hadronic states ($A_1-$meson, say) as well,
and the spectral density $\rho(s_1,s_2,q^2)$ contains also the
part $\rho^{higher \, states}(s_1,s_2,q^2)$ related to other
elastic and transition form factors. This is the price for
 going off the pion mass shell. The problem now is to pick out
the $F_\pi$ term from the whole mess. 

\begin{figure}[t]
\mbox{
   \epsfxsize=12cm
 \epsfysize=3cm
\hspace{-0.5cm} 
  \epsffile{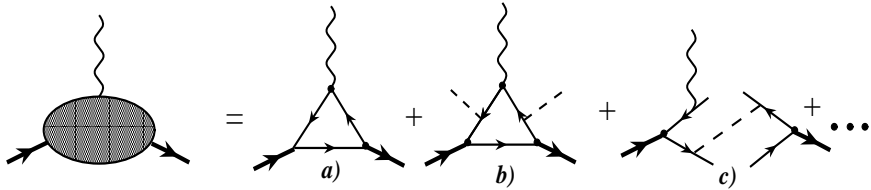}  }
{\caption{\label{fig:ope} 
 Structure of the operator product expansion for $T(p_1^2,p_2^2,q^2)$
 }}
\end{figure}

Of course, calculating $T(p_1^2,p_2^2,q^2)$
in the lowest orders of perturbation theory one never observes
something like the pion pole: one obtains a smooth function
$\rho^{pert}(s_1,s_2,q^2)$ corresponding to transitions between the free-quark
$\bar{u}d-$states with invariant masses $s_1$ and $s_2,$ respectively. The
difference between ``exact'' density $\rho(s_1,s_2,q^2)$ and its perturbative
analog $\rho^{pert}(s_1,s_2,q^2)$ is reflected by additional nonperturbative
contributions to $T(p_1^2,p_2^2,q^2).$
 These contributions are due to quark
and gluon condensates $\langle\bar qq\rangle,$ $\langle GG\rangle$
{etc.}, describing (and/or parameterizing) the nontrivial structure of
the QCD vacuum state.
Formally, these terms appear from the operator product expansion
for the amplitude $T(p_1^2,p_2^2,q^2)$ (Fig.\ref{fig:ope}):
\be
T(p_1^2,p_2^2,q^2)=T^{pert}(p_1^2,p_2^2,q^2)+a\frac{\langle GG\rangle}{(p^2)^3}
 +b\frac{\alpha_s\langle\bar qq\rangle^2}{(p^2)^4}+\ldots\ . \label{eq:ope}
\end{equation}

  The problem now is to construct such a model of the spectral
density $\rho(s_1,s_2,q^2)$ which gives the best agreement between two
expressions for $T$, i.e.,  Eqs. (\ref{eq:dr}) and (\ref{eq:ope}). 
 Naturally, having only a few
first terms of the $1/p^2-$expansion one can hope to reproduce only
the gross features of the hadronic spectrum in the relevant
channel. Still, just using the simple fact that the condensate
contributions die out for large $p^2,$ one obtains the {\em global duality}
relation between quark and hadronic densities
\be
\int_0^\infty ds_1\int_0^\infty ds_2
\left(\rho(s_1,s_2,q^2)-\rho^{pert}(s_1,s_2,q^2)\right)=0. \label{eq:dual}
\end{equation}
In other words,   integrally the two densities are rather similar. One can 
even expect that they are very close for high $s$ to secure the convergence 
of the integral (\ref{eq:dual}). 

The standard ansatz is to
approximate the higher states contribution into $\rho(s_1,s_2,q^2)$ by the
free quark density (compare to  the quasiclassical approximation hor high 
levels in quantum mechanics):
\be
\rho^{higher \, states}(s_1,s_2,q^2)=\left[1-\theta(s_1<s_0)\theta(s_2<s_0)
\right]\rho^{pert}(s_1,s_2,q^2) ,  \label{eq:highst}
\end{equation}
with $s_0$ being the effective threshold for the higher states
production. 

\subsection {Summary} 

To summarize, the basic ingredients of the
QCD sum rule approach are
\\ $\bullet$  Operator product expansion, or factorization of the
short-distance, perturbatively calculable amplitudes, with the 
long-distance information accumulated by the quark and gluon 
condensates --  the basic parameters of the approach -- characterizing
the properties of the QCD vacuum.
\\ $\bullet$  Dispersion relations, relating the perturbatively calculated 
amplitude with the physical hadronic parameters.
\\ $\bullet$  To extract the hadronic parameters one should construct
an ansatz for the hadronic spectrum (like ``first resonance plus
continuum'', { etc.}) and then perform fitting procedure.

Just from the above list it is clear that QCD sum rules cannot
produce exact results. Their precision is unavoidably limited by the 
following facts :
\\ $\bullet$  One can calculate normally only few first terms of the OPE.
\\ $\bullet$  The model ansatz reproduces the real density only approximately.
\\ $\bullet$  Furthermore, in some channels could appear nonperturbative
short-distance contributions (``direct instantons'').

As a result, using the QCD sum rules, 
one can normally expect only 10\% - 20\%  accuracy.

\section{ QCD Sum Rules in Quantum Mechanics}

 We are going to apply the QCD sum rule method in the theory, where
no exact results for hadrons were obtained so far. But,
as formulated above, the QCD sum rule method is quite 
general. In fact, many features of the method can be
demonstrated on the exactly   solvable model of the quantum-mechanical 
oscillator.  In the discussion below,
we  basically follow the pioneering paper\cite{srqm},
with some simplifying modifications\cite{fosc,hab}.
 
\subsection{ Sum rules for the lowest state of  two-dimensional 
oscillator.} 

 Let us consider the simplest example of a system
with confinement, the oscillator, with the confining potential being
\be
V(\vec{x}) = \frac{m\omega^2}{2}r^2   
\label{eq:V(x)}.
\end{equation}
Formulas are the simplest (without square roots, complicated factorial
terms, {etc.}) if we choose  the oscillator in 
two spatial dimensions ($+$ time dimension).
In this case the energy levels are given by 
\be
E_n = (2n+1)\omega,
\label{eq:En}
\end{equation}
and the values of the wave functions 
 at the origin (quantum-mechanical analogues
of $f_{\pi}$ and other leptonic decay parameters) are the same for all 
levels corresponding to
radial excitations with  zero angular momentum:
\be
|\psi_n(0)|^2 = \frac{m\omega}{\pi}.
\label{eq:psinull}
\end{equation}

To calculate  $E_0$ and $\psi_0(0)$ 
using the sum rule approach one should proceed in a rather peculiar way: 
instead of concentrating on the lowest level, one should 
consider the weighted sum of all levels:
\be
M(\epsilon) = \sum_{k=0}^{\infty} \  |\psi_k(0)|^2 e^{-E_k/{\epsilon}} 
\label{eq:Meps}
\end{equation}
in which the lowest level dominates only for very small values of
$\epsilon$. For large $\epsilon$ all levels are essential.
This combination (\ref{eq:Meps}) is directly related to the two-time
Green function 
\be
G(x_1,t_1 | x_2,t_2) = \sum_{k=0}^{\infty} \psi_k^*(x_2) \psi_k(x_1)
e^{-iE_k(t_2-t_1)}
\label{eq:Gxt}
\end{equation}
describing the probability amplitude for 
transition from the point $x_1$ at time $t_1$ to the point $x_2$ at
time $t_2$. To get $M(\epsilon)$ one should take $x_2=x_1=0, \, t_1=0$ 
and $t_2=1/{i\epsilon}$. In other words, the transition is 
taken in the imaginary time.

In our case, explicit form of $M(\epsilon)$ can be easily calculated
using Eqs. (\ref{eq:En}) and (\ref{eq:psinull}):
\be
M^{osc}(\epsilon)  =  \sum_{k=0}^{\infty} \left (\frac{m\omega}{\pi}
\right ) 
e^{-(2k+1)\omega/{\epsilon}}= 
\frac{m\omega}{\pi \sinh(\omega/\epsilon)}
\label{eq:Mosc}
\end{equation}
In the large-$\epsilon$ limit one has
\be
M^{osc}(\epsilon) = \frac{m\epsilon}{2\pi} \ \left (1-\frac{\omega^2}{6\epsilon^2} +
\frac7{360}\frac{\omega^4}{\epsilon^4}-\frac{31}{15120}
\frac{\omega^6}{\epsilon^6}+\ldots \right )
\label{eq:Mexpand}
\end{equation} 
{\em i.e.,} the first term of the large-$\epsilon$ ($\equiv$small-time) 
expansion has no dependence
on $\omega$, the paramater characterizing the strength of oscillator 
 potential.  
This simply means that at small time differences the Green function
coincides with the free one:
\be
G^{free}(2|1)=\frac{m}{2\pi i(t_2-t_1)} \exp \left \{-i\frac{m(x_2-x_1)^2}
{2(t_2-t_1)} \right \} 
\label{eq:Gfree}
\end{equation}
and
\be
M^{free}(\epsilon) = \frac{m\epsilon}{2\pi}
\label{eq:Mfree}
\end{equation}
The corrections to the free-particle behavior in Eq. (\ref{eq:Mexpand}) 
are in powers of $\omega^2$. This corresponds to the perturbation expansion
in the potential $V(r)$ (\ref{eq:V(x)}):
\be
M= M^{free}+ M^{free} \otimes V \otimes M^{free} + 
M^{free} \otimes V \otimes M^{free} \otimes V \otimes M^{free}
+ \ldots
\label{eq:MMfree}
\end{equation}

The function $M^{free}(\epsilon)$ can also be represented as a sum over
the (free)  states with $E_k=k^2/{2m}$:
\be
M^{free}(\epsilon) = \int \frac{d^2k}{(2\pi)^2}e^{-E_k/\epsilon} = 
\frac{m}{2\pi} \int_0^{\infty} e^{-E/\epsilon} dE \equiv \frac1{\pi}
\int_0^{\infty} e^{-E/\epsilon} \rho(E) dE \  .
\label{eq:Mfreespectral}
\end{equation}
We introduced here a  new function
\be
\rho^{free}(E) = \frac{m}{2}\, \theta(E>0),
\label{eq:rhofree}
\end{equation}
the spectral density corresponding to the free motion, which will play
an important role in the subsequent discussion. The oscillator
function $M^{osc}(\epsilon)$ also can be written 
in the spectral representation with
the spectral density $\rho^{osc}(E)$ which is in this case
 a superposition of the delta-functions:
\be
\rho^{osc}(E) = m\omega \sum_{k=0}^{\infty} \delta(E-(2k+1)\omega).
\label{eq:rho}
\end{equation}
At first sight, the two densities are completely different.
However, if one integrates them from zero to the midpoint between the
first and second levels, one obtains the same result:
\be
\frac1{m}\int_{0}^{2\omega} \rho^{free}(E) dE = \omega =
 \frac1{m} \int_{0}^{2\omega} \rho^{osc}(E) dE  \  .
\label{eq:dual0}
\end{equation}
Moreover, similar relations hold for other levels:
\be
\int_{2k\omega}^{2(k+1)\omega} \rho^{free}(E) dE = m\omega =
 \int_{2k\omega}^{2(k+1)\omega} \rho^{osc}(E) dE  \  .
\label{eq:dualk}
\end{equation}
The equality is not violated even if one multiplies $\rho(E)$
by an extra power of the energy $E$:
\be
\int_{2k\omega}^{2(k+1)\omega} \rho^{free}(E) \, E dE = m\omega^2(2k+1) =
 \int_{2k\omega}^{2(k+1)\omega} \rho^{osc}(E) \, E dE \  .
\label{eq:dualkE}
\end{equation}
Thus, there exists the {\em local duality} between each resonance
and the free states. It looks like each level ``takes'' its part
of the free spectral density, leaving the integral unchanged.

Turning back to the $\omega$-expansion for $M^{osc}(\epsilon)$ 
(Eq. (\ref{eq:Mexpand}))
\be
M^{osc}(\epsilon) =  M^{free}(\epsilon) \ \left
\{1-\frac{\omega^2}{6\epsilon^2} +
\frac7{360}\frac{\omega^4}{\epsilon^4}-\frac{31}{15120}
\frac{\omega^6}{\epsilon^6}+\ldots \right \}
\end{equation}
one can write
\be
M^{osc}(\epsilon) - M^{free}(\epsilon) = O( \omega^2 / \epsilon) +
O(\omega^4 / \epsilon^3) + \ldots \, .
\label{eq:M-M}
\end{equation}
or, in terms of the spectral densities
\be
\frac1{\pi} \int_0^{\infty} \left 
[\rho^{osc}(E) - \rho^{free}(E) \right ] e^{-E/\epsilon} dE
= \sum_{k=1} \frac{A_k}{\epsilon^k}.
\label{eq:rho-rho}
\end{equation}
The terms on the rhs, suppressed for large $\epsilon$ 
by powers of $1/\epsilon$ 
 are normally referred to as ``power corrections''. Thus, the difference
between the integrated spectral densities vanishes as $\epsilon \to \infty$:
\be
\frac1{\pi} \int_0^{\infty} [\rho^{osc}(E) - \rho^{free}(E)] dE =0 \, , 
\label{eq:globald}
\end{equation}
since there is no $O(1)$ term in the rhs of Eq. (\ref{eq:rho-rho}).
Eq. (\ref{eq:globald}) is a standard {\em global duality} relation. In quantum
mechanics one can obtain such a relation for any potential regular at
the origin.

The lesson is that the two densities, despite the apparent absence of 
similarity, are rather close to each other in an integral sense.
Now, the idea is to incorporate this similarity in the sum rule
\be
|\psi_0(0)|^2 e^{-E_0/\epsilon} + ``higher \  states" =
\frac{m \epsilon}{2 \pi } \left  \{1-\frac{\omega^2}{6\epsilon^2} +
\frac7{360}\frac{\omega^4}{\epsilon^4}-\frac{31}{15120}
\frac{\omega^6}{\epsilon^6}+\ldots \right \} 
\label{eq:oscsr1}
\end{equation}
by substituting the ``higher states'' contribution by that of the
free states having energy above some threshold $s_0$.
The sum rule then reads
\be
|\tilde{\psi}_0(0)|^2 e^{-E_0/\epsilon} \approx 
\frac{\epsilon}{2} \, \left (1-e^{-s_0/\epsilon} \right 
) - \frac{\omega^2}{12\epsilon} +
\frac7{720}\frac{\omega^4}{\epsilon^3}-\frac{31}{30240}
\frac{\omega^6}{\epsilon^5}+\ldots 
\label{eq:oscsr2}
\end{equation} 
 where 
$$
|\tilde{\psi}_0(0)|^2 = \frac{\pi}{m}|{\psi}_0(0)|^2
$$
The parameters to be extracted from the sum rule are $E_0, \tilde{\psi}_0(0)$
and $s_0$. The first step is to extract $E_0$. To this end we differentiate
Eq. (\ref{eq:oscsr2}) with respect to $(-1/\epsilon)$ to get
\be
|\tilde{\psi}_0(0)|^2 E_0 e^{-E_0/\epsilon} \approx 
\frac{\epsilon^2}{2} -\frac{\epsilon^2}{2} \left
 (1+\frac{s_0}{\epsilon} \right )
e^{-s_0/\epsilon} +  \frac{\omega^2}{12} -
\frac7{240}\frac{\omega^4}{\epsilon^2}+\frac{31}{6048}
\frac{\omega^6}{\epsilon^4}+\ldots 
\label{eq:oscsr2diff}
\end{equation}
The ratio of the two equations just gives the expression for $E_0$:
\be
E_0 \approx  \frac{\frac{\epsilon^2}{2} -\frac{\epsilon^2}{2} (1+\frac{s_0}{\epsilon})
e^{-s_0/\epsilon} +  \frac{\omega^2}{12} -
\frac7{240}\frac{\omega^4}{\epsilon^2}+\frac{31}{6048}
\frac{\omega^6}{\epsilon^4}+\ldots}
{\frac{\epsilon}{2} (1-e^{-s_0/\epsilon}) - \frac{\omega^2}{12\epsilon} +
\frac7{720}\frac{\omega^4}{\epsilon^3}-\frac{31}{30240}
\frac{\omega^6}{\epsilon^5}+\ldots}  \  .
\label{eq:E0}
\end{equation}

If we would take the whole series in $\omega/\epsilon$ and the exact 
expression for the higher states contribution, then  the result
for $E_0$ would be the $\epsilon$-independent  exact value $E_0 = \omega$.
However, under approximations we made, the $E_0$-value 
extracted from Eq. (\ref{eq:E0}) depends both on the auxiliary 
parameter $\epsilon$ and the threshold parameter $s_0$.
The standard strategy is to take the value of $s_0$, for which
$E_0$ has the minimal dependence on $\epsilon$ in the 
intermediate region $1< \epsilon/\omega <2$ where both the higher power
corrections and the error due to the rough model of the higher states
contribution are small ($ \approx 5 - 10 \%$). Truncating the series
by terms of the second order in $\omega^2$, one obtains\footnote{Plotting
$E_0$ for various values of $s_0$ is absolutely 
straightforward,  and we encourage the readers 
perform the fitting as an excercise.}  from
such a fitting $s_0 = 1.6 \omega$ and $E_0 = 0.9 \omega$. 
Adding the third order terms improves the result:
$s_0 = 1.75 \omega$ and $E_0 = 0.95 \omega$. 
In our case one can substitute into the sum rule the all-order result.
The remaining $\epsilon$-dependence is then entirely due to the use
of an approximate model for the higher states. For sufficiently small
values of $\epsilon$ all the curves give the exact value $E_0=2\omega$.
The flattest curves are those with $s_0 \sim 2\omega$.

The second step is to extract the $\tilde{\psi}(0)$ value from the 
original sum rule (\ref{eq:oscsr2})
\be
|\tilde{\psi}_0(0)|^2 \approx  e^{E_0/\epsilon} \left \{\frac{\epsilon}{2}
(1-e^{-s_0/\epsilon}) - \frac{\omega^2}{12\epsilon} +
\frac7{720}\frac{\omega^4}{\epsilon^3}-\frac{31}{30240}
\frac{\omega^6}{\epsilon^5}+\ldots \right \} 
\label{eq:oscsr3} \, . 
\end{equation}
The analysis goes in a similar way. If one takes the third-order value
$E_0 = 0.95\, \omega$, one obtains $|\tilde{\psi}_0(0)|^2 \sim 0.9\, \omega$.
In the $\epsilon \to \infty$ limit one obtains the local duality relation
\be
\{|\tilde{\psi}_0(0)|^2\}^{LD} = \frac{s_0}{2}.
\label{eq:psi0ld}
\end{equation}
So, the ``exact value'' for the duality interval $s_0$ is $2\omega$.

Summarizing, using the sum rule method one can determine the parameters
of the lowest resonance with $ \approx 10 \%$ accuracy. The main source
of the error is the use of a rather rough model for the spectral density
of the higher states. It should be noted that in QCD 
situation is better: normally only the first resonance is narrow, 
while higher states are broader and broader, so that approximating them 
by  free quark functions does not lead to large errors.

\subsection{ Sum rules for form factors of the oscillator states.} 

To calculate the form factors
\be
F_{kl}(Q) = \int  \psi^*_k(x) \psi_l(x) e^{iQx} \,d^2x 
\label{eq:Fkl}
\end{equation}
within the sum rule approach, one should construct a similar integral
using the Green functions (\ref{eq:Gxt})
\be
G(\,x,\, 1/ i\epsilon \, |\,0,\,0\,) = 
\sum_{k=0}^{\infty} \psi_k^*(x_2) \psi_k(x_1)
e^{-E_k/\epsilon} \, .
\label{eq:Gxeps}
\end{equation}
This gives the amplitude
\ba
M(\epsilon_1,\epsilon_2,Q^2) 
& = & \int G(\,x,\, 1/ i\epsilon_1 \,|\,0,\,0\,) \, 
G^*(\,x,\, 1/{i\epsilon_2}\,|\,0,\,0\,) e^{iQx}\,d^2x   \nonumber \\
& = & \sum_{k,l=0}^{\infty} \psi^*_l(0)\psi_k(0) 
\left [\int e^{iQx} \psi^*_k(x) \psi_l(x) d^2x \right ] 
e^{-E_k/{\epsilon_1} - E_l/{\epsilon_2}}  \nonumber \\ 
& = & \sum_{k,l=0}^{\infty}  \psi^*_l(0)\psi_k(0) F_{kl}(Q^2)
e^{-E_k/{\epsilon_1} - E_l/{\epsilon_2}}
\label{eq:Mepsq}
\ea
One can write this amplitude in the (double) spectral representation:
\be
M(\epsilon_1,\epsilon_2,Q^2) = \frac1{\pi^2} \int_0^{\infty} 
\int_0^{\infty}  \,e^{-s_1/\epsilon_1 - s_2/\epsilon_2} 
\rho(s_1,s_2,Q^2) ds_1 ds_2
\label{eq:M2rho}
\end{equation}
where the spectral function is the sum of double delta-functions:
\be
\rho(s_1,s_2,Q^2) = \pi^2 \sum_{k,l=0}^{\infty} \psi^*_l(0)\psi_k(0) \,
 F_{kl}(Q)\, \delta(s_1-E_k)\, \delta (s_2-E_l).
\label{eq:rhoff}
\end{equation}
In the free case one has
\be
G^{free}(\,x,\, 1/i\epsilon \,|\,0,\,0)=\frac{m\epsilon}{2\pi} 
e^{-mx^2\epsilon/2} =
\int \frac{d^2k}{(2\pi)^2} e^{ikx}e^{-k^2/{2m\epsilon}}
\label{eq:Gfreek}
\end{equation}
for the Green function and
\ba
M^{free}(\epsilon_1,\epsilon_2,Q^2) 
& = & \int \frac{d^2k}{(2\pi)^2} e^{-k^2/{2m\epsilon_1}-(k+Q)^2/{2m\epsilon_2}}
\nonumber \\
& = & \left (\frac{m}{2\pi}\right )
\frac{\epsilon_1\epsilon_2}{\epsilon_1+\epsilon_2}
\exp{\left (-\frac{Q^2}{2m(\epsilon_1+\epsilon_2)}\right )}
\label{eq:Mfreek}
\ea
for the amplitude. The free spectral function $\rho^{free}(s_1,s_2,Q^2)$ can
be easily extracted:
\ba
\rho^{free}(s_1,s_2,Q^2)
& = & \pi^2 \int \frac{d^2k}{(2\pi)^2}\,
 \delta \left (s_1-\frac{k^2}{2m} \right )
\delta \left (s_2-\frac{(k+Q)^2}{2m} \right )  \nonumber \\
\vspace{0.5cm}
& = & \frac{\theta (|\sqrt{2ms_1}+ \sqrt{2ms_2}| \geq |Q| \geq |\sqrt{2ms_1}-\sqrt{2ms_2}|)}
{\sqrt{4s_1s_2-(s_1+s_2-Q^2/{2m})^2}}.
\label{eq:rho2free}
\ea
The restriction on $s_1, s_2$ is in fact the triangle relation 
$$|k_1|+|k_2|\geq |Q| \geq |\,|k_1|-|k_2|\,|$$ 
between the initial momentum $k_1$ ($|k_1|=\sqrt{2ms_1})$, 
the momentum transfer $Q$
and the final momentum $k_2 = (k_1 + Q)$ ($|k_2|=\sqrt{2ms_2})$.

The oscillator Green function is also known:
\be
G^{osc}(\,x,\, 1/ i\epsilon \,|\,0,\,0)= 
\frac{m\omega}{2\pi\sinh(\omega/\epsilon)}
\exp \left
(-\frac{m\omega\vec{x}^2 \cosh (\omega/\epsilon)}
{2\sinh(\omega/\epsilon)}\right )
\label{eq:Gosc}
\end{equation}
and using it, one can calculate the amplitude
\be
M(\epsilon_1,\epsilon_2,Q^2) = \frac{m\omega}
{2\pi\sinh(\omega/\epsilon_1 +\omega/\epsilon_2)}
\exp \left (-\frac{Q^2}{2m\omega} 
\frac{\sinh({\omega}/{\epsilon_1})\sinh({\omega}/{\epsilon_2})}
{\sinh({\omega}/{\epsilon_1}+{\omega}/{\epsilon_1})} \right ) \  .
\label{eq:M2osc}
\end{equation}
Now, representing $M$ as
\ba
M(\epsilon_1,\epsilon_2,Q^2) & = & \frac{m\omega}{\pi}  
\sum_{k=0}^{\infty} 
e^{-(2k+1)\omega(1/{\epsilon_1}+1/{\epsilon_2})}  
\exp{\left (-\frac{Q^2}{4m\omega} \right )} \nonumber \\
& & 
 \exp{ \left ( \frac{Q^2}{4m\omega}
\left [1- \frac{(1-e^{-\omega/{\epsilon_1}})(1-e^{-\omega/{\epsilon_2}})}
{(1-e^{-\omega/{\epsilon_1}}e^{-\omega/{\epsilon_2}})} \right ]  \right )}
\ea
and expanding $M$ in powers of $e^{-\omega/\epsilon_1} \equiv {\cal E}_1$
and $e^{-\omega/\epsilon_2} \equiv {\cal E}_2$ 
one can extract the form factors $F_{lk}(Q^2)$:
\ba
M(\epsilon_1,& \epsilon_2, & Q^2)  =  \frac{m\omega}{\pi}
\sum_{k=0}^{\infty} {\cal E}_1^{2k+1} {\cal E}_2^{2k+1}
\exp{\left (-\frac{Q^2}{4m\omega} \right )} \nonumber \\
 & & \hspace{-2cm}\left \{ 1 + \frac{Q^2}{4m\omega} \left 
[1-\frac{(1-{\cal E}_1)(1-{\cal E}_2)}
{1- {\cal E}_1 {\cal E}_2} \right ] + \frac1{2} \left 
(\frac{Q^2}{4m\omega} \right )^2 [ \ldots ]^2 
+ \ldots \right \}
\ea
identifying them as the factors in front 
of the ${\cal E}_1^k {\cal E}_2^l$ terms:
\ba
F_{00}(Q^2) = \exp{\left (-\frac{Q^2}{4m\omega} \right )} \  ,  \nonumber \\
F_{01}(Q^2) = F_{10}(Q^2) = \frac{Q^2}{4m\omega} 
\exp{\left (-\frac{Q^2}{4m\omega}  \right )} \ ,  \nonumber \\
F_{11}(Q^2) = \left (1-\frac{Q^2}{4m\omega} \right )^2 
\exp{\left (-\frac{Q^2}{4m\omega} \right )} 
\ea

It is instructive to write down the amplitude 
$M(\epsilon_1, \epsilon_2,  Q^2)$ for $Q^2 = 0$:
\be
M(\epsilon_1, \epsilon_2,  Q^2=0) =   \frac{m\omega}
{2\pi\sinh(\omega/\epsilon_1 +\omega/\epsilon_2)}
=  \frac{m\omega}{\pi}  
\sum_{k=0}^{\infty} {\cal E}_1^{2k+1} {\cal E}_2^{2k+1} \ .
\end{equation}
It is easy to notice that
\\ $\bullet$  There are no terms with $k\neq l$: only the diagonal transitions
are allowed in this limit
\\ $\bullet$  The numerical factor is just the same as in Eq. (\ref{eq:Mosc}) 
for $M^{osc}(\epsilon)$. 
This means
that $F_{kk}(0) = 1 $ for all $k$, just as required by charge 
conservation.

Next step is to get the perturbative expansion in powers of $\omega$ :
\ba
M(\epsilon_1, \epsilon_2, Q^2)
& = & \frac{m\epsilon_1 \epsilon_2}{2\pi(\epsilon_1 + \epsilon_2)}
 \ \left \{ \left 
[1-\frac{\omega^2}{6} \left 
(\frac1{\epsilon_1}+\frac1{\epsilon_2}\right )^2 + \ldots \right ] \right.
\nonumber \\
&   &  \left. \times \exp \left (-\frac{Q^2}{2m}
\frac{\frac1{\epsilon_1} (1+ \frac1{6}\frac{\omega^2}{\epsilon_1^2} + \ldots) 
\frac1{\epsilon_2} (1+ \frac1{6}\frac{\omega^2}{\epsilon_2^2} + \ldots)}
{ (\frac1{\epsilon_1}+\frac1{\epsilon_2})(1+ \frac{\omega^2}{6} 
(\frac1{\epsilon_1}+\frac1{\epsilon_2})^2 + \ldots)} \right ) 
\right \} \nonumber \\
& = &  \frac{m\epsilon_1 \epsilon_2}{2\pi(\epsilon_1 + \epsilon_2)}
\ \left \{ \left [1-\frac{\omega^2}{6}
\left (\frac1{\epsilon_1}+\frac1{\epsilon_2} \right )^2 +
\frac{7\omega^4}{360} \left (\frac1{\epsilon_1}+
\frac1{\epsilon_2} \right )^4 + \ldots \right ] \right. \nonumber \\
&   & \hspace{2cm} \left. \exp{ \left (-\frac{Q^2}
{2m(\epsilon_1+\epsilon_2)} \right )} 
\left [1+\frac{Q^2\omega^2}{6m\epsilon_1\epsilon_2(\epsilon_1+\epsilon_2)} 
\right. \right. \nonumber \\ & - & \left. \left. 
\frac{Q^2\omega^4[(\epsilon_1+\epsilon_2)^2+2\epsilon_1\epsilon_2]}
{90m\epsilon_1^3\epsilon_2^3(\epsilon_1+\epsilon_2)} +
\frac{Q^4\omega^4}{72m^2 \epsilon_1^2\epsilon_2^2(\epsilon_1+\epsilon_2)^2}
+ \ldots \right ] \right \} 
\nonumber \\
&  & \hspace{1cm} = M^{free}(\epsilon_1, \epsilon_2, Q^2) 
\left \{1+ \sum_{k=1}^{\infty}
(\omega^2)^k \varphi_k(\epsilon_1, \epsilon_2, Q^2) \right \}
\label{eq:M2expand}
\ea

The sum rule now reads
\ba
|\psi_0(0)|^2 F_{00}(Q^2)  
e^{-E_0/\epsilon_1-E_0/\epsilon_2} & + &``higher \, states" 
\nonumber \\ &  & \hspace{-3cm} = M^{free}(\epsilon_1, \epsilon_2, Q^2) 
 \left \{1+ O\left (\frac{\omega^2}{\epsilon^2} \right )+
 O\left (\frac{\omega^4}{\epsilon^4} \right )+ \ldots \right \}
\ea

To proceed further, one should take some ansatz 
for the higher states contribution. 
The ansatz similar to that used earlier is to assume that 
$$ ``higher \, states" = ``free \, states" $$
outside the square $(0,s_0)\otimes (0,s_0)$.
Then the sum rule has the form
\ba
|\psi_0(0)|^2 F_{00}(Q^2) e^{-E_0/\epsilon_1-E_0/\epsilon_2} \nonumber \\
&  &\hspace{-2cm}= \frac1{\pi^2} \int_0^{s_0} ds_1 
\int_0^{s_0} ds_2 \, \rho^{free}(s_1,s_2,Q^2)
e^{-s_1/\epsilon_1-s_2/\epsilon_2}  \nonumber \\ 
&  &  + O \left (\frac{\omega^2}{\epsilon} \right )  
 + O \left (\frac{\omega^4}{\epsilon^3} \right ) + \ldots  \ .
\ea
In the $\epsilon_1,\epsilon_2 \to \infty$ limit one obtains the 
local duality relation:
\be
|\psi_0(0)|^2 F_{00}(Q^2) = \frac1{\pi^2} \int_0^{s_0} 
\int_0^{s_0} \, \rho^{free}(s_1,s_2,Q^2) ds_1 ds_2
\end{equation}
with $\rho^{free}(s_1,s_2,Q^2) $ given by
Eq. (\ref{eq:rho2free}):
\be
\rho^{free}(s_1,s_2,Q^2) =  \frac1{4}\int d^2k \,
 \delta \left (s_1-\frac{k^2}{2m} \right )\delta 
\left (s_2-\frac{(k+Q)^2}{2m} \right ).
\end{equation}
Using the two expressions given above, one can write
\be
|\psi_0(0)|^2 F_{00}^{LD}(Q^2) =  \frac1{(2\pi)^2} \int d^2k \,
 \theta \left (\frac{k^2}{2m}<s_0\right )
\theta \left (\frac{(k+Q)^2}{2m}<s_0 \right ).
\end{equation}
This integral can be easily calculated if one notices that it 
defines the overlap area of two circles having the same radius 
$R^2=2ms_0$ , with their 
 centers separated by the distance equal to $Q$:
\be
|\psi_0(0)|^2 F_{00}^{LD}(Q^2)= 
\frac{ms_0}{\pi^2} 
\left [\, \cos^{-1} (\kappa )-
\kappa
\sqrt{1-\kappa^2}\, \right ]\, \theta(Q^2<8ms_0) \  , 
\end{equation}
where $\kappa = {Q}/{\sqrt{8ms_0}}$.
 Recall now that from the local duality relation for $M(\epsilon)$
we had
\be
\{|\psi_0(0)|^2\}^{LD} = \frac{ms_0}{2\pi}
\end{equation}
and this gives the result
\be
F_{00}^{LD}(Q^2) = 
\frac{2}{\pi}\, \left [\, \cos^{-1} (\kappa )-
\kappa
\sqrt{1-\kappa^2}\, \right ]\,\theta(Q^2<8ms_0)  
\end{equation}
having the correct normalization
$ F_{00}^{LD}(Q^2) = 1 $.

The local duality prediction calculated for the ``exact value''
$s_0=2\omega$ of the duality interval
\ba
F_{00}^{LD}(Q^2)|_{s_0=2\omega} = 
\frac{2}{\pi} \left [\cos^{-1} \left (\frac{Q}{\sqrt{16m\omega}} \right )
\right. \nonumber \\  \left.
-\frac{Q}{\sqrt{16m\omega}}
\sqrt{1-\frac{Q^2}{16m\omega}}\, \right ]\,\theta(Q^2<16m\omega)
\ea
is  very close to  the exact form factor
\be
F_{00}^{exact}(Q^2) = \exp{\left (-\frac{Q^2}{4m\omega} \right )} \  .
\end{equation}

There is a very simple physics behind the local duality. One should
just compare the local duality prescription for the form factor
\be
 F_{00}^{LD}(Q^2)
 =\frac1{|\psi_0(0)|^2}   \int \frac{d^2k}{(2\pi)^2} \,
 \theta \left (\frac{k^2}{2m}<s_0 \right )\,
\theta \left (\frac{(k+Q)^2}{2m}<s_0 \right )
\ee
and the exact formula
\be
 F_{00}(Q^2)
 = \int \frac{d^2k}{(2\pi)^2} \,
 \psi_0(k)\psi_0^*(k+Q)
\end{equation}
to realize that the local duality, in fact, is equivalent to using
a model form for the wave function in the momentum space:
\be
\tilde{\psi}_{0}^{LD}(k)
 =\frac1{|\psi_0(0)|} \, \theta(k^2<2ms_0)
\end{equation}
or, taking $s_0=2\omega$:
\be
\tilde{\psi}_{0}^{LD}(k)
 =\sqrt{\frac{\pi}{m\omega}} \, \theta(k^2<4m\omega),
\end{equation}
to be compared with the exact wave function
\be
\tilde{\psi}_{0}(k)  =2\sqrt{\frac{\pi}{m\omega}} 
\exp \left (-\frac{k^2}{2m\omega} \right ).
\end{equation}
It is easy to establish that the model wave function correctly 
reproduces the simplest integral properties of the exact wave function:
\ba
\int \tilde{\psi}^{exact}(k)d^2k = \int \tilde{\psi}^{LD}(k) d^2k \\
\int |\tilde{\psi}^{exact}(k)|^2 d^2k = \int |\tilde{\psi}^{LD}(k)|^2 d^2k.
\ea

\section{ Condensates ant Operator Product Expansion} 

\subsection{ Correlators }

 The basic object in the quantum mechanical example was the Green 
function 
$G(\,x_1,\,t_1\, |\,x_2,\,t_2\,)$ 
describing the propagation of a particle from one point to another.
In fact, it was sufficient to analyze the probability amplitude 
for the partricle to 
remain at the same spatial point after some 
imaginary time elapsed: 
\begin{equation}
M(\epsilon) =  G(\,0,\,0\,|\,0,\, 1/{i\epsilon}\,) \ .
\label{eq:Geps}
\end{equation}

The question is what should be its analog in the quantum field theory? 
Recall, that the motion in the external potential corresponds to a 
two-particle system with one  of the particles being infinitely massive.
If this particle moves, the other moves together with it. 
So,  the  Green function describes really a collective motion of two particles
from one point to another. In quantum field theory this corresponds to
the Green-like functions 
$$ 
\langle 0 | T \,(j(x)\,j(y)\,)|\,0\rangle
$$
normally referred to as correlators, with the currents $j(x)$ being 
the products of fields at the same point, {\em e.g.} 
$J^\mu(x)=\bar q(x)\gamma^\mu q(x)$ .  Due to the translation invariance
$$ 
\langle 0 | T \,(j(x)\,j(y)\,)|\,0\rangle \to 
\langle 0 | T \,(j(x-y)\,j(0)\,)|\,0\rangle
$$
it is always possible and convenient to take one point at the origin.
In QFT, it is also more convenient to consider correlators in the
momentum representation:
$$
\Pi(q) = \int e^{iqx}  \langle 0 | T \,(j(x)\,j(0)\,)|\,0\rangle\, d^4 x.
$$
 In the lowest order the correlator is given by the product of the
relevant propagators, but there are also radiative corrections, which 
one can calculate in perturbative QCD, at least in the region where
$q$ is spacelike $q^2<0$ and large enough.

Next step is to relate $\Pi(q)$ to a sum over the physical states. 
This is performed {\em via} the dispersion relation
$$
\Pi(q^2) = \frac1{\pi} \int_0^{\infty} \frac{\rho(s)}{s-q^2} ds + 
``subtractions".
$$
The subtraction terms  
appear if the spectral density is such that
the dispersion integral diverges. If, say, $\rho(s) \sim s^n$, one
needs $(n+1)$ subtractions:
\ba
\Pi(q^2) & = & \frac1{\pi}\int_0^{\infty} \frac{\rho(s)}{s-q^2} ds \nonumber \\
& - & \frac1{\pi} \int_0^{\infty}ds\, \rho(s) \biggl \{ \frac1{s-\lambda^2} +
\frac{q^2-\lambda^2}{(s-\lambda^2)^2} +  \ldots +
\frac{(q^2-\lambda^2)^n}{(s-\lambda^2)^{n+1}}\}  \nonumber \\
 & + & \{ \Pi(\lambda^2)+ (q^2-\lambda^2)\Pi^{\prime}(\lambda^2) + \dots +
\frac{(q^2-\lambda^2)^n}{n!}\Pi^{(n)}(\lambda^2) \biggr \}. 
\ea
After collecting all the terms containing $\rho(s)$, one observes that $\rho(s)$
is divided by $(s-q^2)(s-\lambda^2)^{n+1}$, and the integral converges.
Note, that all the subtraction terms are polynomials of a
finite order in $Q^2$ (from now on  we use the notation 
$ q^2 = - Q^2$,
with $Q^2>0$). 

\subsection{ Borel transformation}

 Note, that 
in the oscillator studies, we dealt with the exponentially weighted sum:
\be
M(\epsilon) = \frac1{\pi} 
\int_0^{\infty}  \,e^{-s/\epsilon} 
\rho^{osc}(s) \, ds.
\label{eq:Mrho}
\end{equation}
We observed then that it has a nice property that for 
$\epsilon$ values of an order of $\omega$, both
the contributions of the higher states were strongly damped 
 and the perturbative 
expansion  for $M(\epsilon)$ was converging fast enough. With the 
power weight implied by the dispersion relation, the suppression 
of the higher states contribution is not so effective. 

However, it is
rather easy to transform the original dispersion integral
into the exponentially weighted one.
To this end, one should apply the so-called Borel transformation,
formally defined by the formula\cite{SVZ} 
\be
\Phi(M^2) =  \lim_{\stackrel{Q^2,n \to \infty} {Q^2/n=M^2}} \frac1{(n-1)!}
(Q^2)^n \left [-\frac{d}{dQ^2} \right ]^n \Pi(Q^2).
\label{eq:fi} 
\end{equation}
 It is easy to check that applying the above procedure to $1/(s+Q^2)$
one really obtains an exponential. Indeed,
\be
\left [-\frac{d}{dQ^2}\right ]^n \frac1{s+Q^2} = \frac{n!}{(s+Q^2)^{n+1}}.
\end{equation}
Multiplying the result by $\frac{(Q^2)^n}{(n-1)!}$ one obtains
\be
n\, \frac{(Q^2)^n}{(s+Q^2)^{n+1}} = \frac1{Q^2/n}\frac1{(1+s/Q^2)^{n+1}} =
\frac1{M^2}\frac1{(1+{s}/{nM^2})^{n+1}}.
\end{equation}
Finally, taking the $n\to\infty$ limit one obtains $\frac1{M^2}e^{-s/M^2}$.
This means that 
\be
\Phi(M^2) \equiv B(Q^2 \to M^2)\, \Pi(Q^2) = \frac1{\pi} 
\int_0^{\infty} \frac{ds}{M^2} \rho(s) e^{-s/M^2} + ``nothing"
\end{equation}
where ``nothing'' means that all the subtraction terms have disappeared,
since no polynomial can survive the  infinite number 
of differentiations.
Thus, the function $\Phi(M^2)$ is the QFT analog of $M(\epsilon)$.

In many cases one can perform the Borel transformation using the 
formula\cite{nucff} 
\be
B(Q^2 \to M^2)\{e^{-AQ^2}\} = \delta (1-AM^2).
\end{equation}
As an illustration, let us reproduce the result obtained above :
\ba
B \frac1{s+Q^2} & = & B \int_0^{\infty} e^{-\alpha(s+Q^2)}\,d\alpha 
\nonumber \\ & = & 
\int_0^{\infty}  e^{-\alpha s} \delta(1-\alpha M^2) \,d\alpha = 
\frac1{M^2}e^{-s/M^2}.
\ea
Other functions, the Borel transforms of which we will need,
are those having a power behavior in $1/Q^2$. In this case we have
\ba
B \frac1{(Q^2)^n} &=& B \int_0^{\infty}
e^{-\alpha Q^2} \frac{\alpha^{n-1}}{(n-1)!}  \,d\alpha \nonumber \\ &=&
\int_0^{\infty}  \delta(1-\alpha M^2) \frac{\alpha^{n-1}}{(n-1)!}\,d\alpha  = 
\frac1{(n-1)! (M^2)^n}.
\ea
Thus, a power expansion for $\Phi(M^2)$ over $1/M^2$ converges 
much faster than the original expansion of $\Pi(Q^2)$ over $1/Q^2$, since
each term is now suppressed  factorially. This looks precisely
like the Borel improvement of a series, and that is why Shifman, Vainshtein 
and Zakharov\cite{SVZ} referred to 
the transformation they introduced  as ``the Borel 
transformation''. The representation for $\Phi(M^2)$ has the form of the
Laplace transformation, and that is why some people call the sum rules
obtained in this way ``the Laplace transform sum rules''. 
In fact, there is no contradiction: $\Phi(M^2)$ is the Borel transform
of $\Pi(Q^2)$ and the Laplace transform of $\rho(s)$. 

To summarize, the Borel transformation produces two improvements:
\\ $\bullet$  higher state contributions in the spectral representation
for $\Phi(M^2)$ are exponentially suppressed,
\\ $\bullet$  its power series expansion in $1/{M^2}$ is factorially improved.

\subsection{ Heavy-light system and the $m \to \infty $ limit}

Using the basic formula, it is rather easy to establish\cite{iw}
 that in the limit
when one of the particles becomes 
infinitely massive, the  QFT function $\Phi(M^2)$
converts into its quantum-mechanical analog. Consider the lowest-order
diagram. The propagator of the heavy particle is
$$
\frac1{(q+k)^2-m^2}= \frac1{q^2-m^2+2qk+k^2} \, .
$$
In the $m \to \infty$ limit all the bound state masses are equal to this mass 
plus some finite quantity. So, it makes sense to subtract this infinite 
constant. To this end one can represent the total momentum $q$ as
$$
q=mv+l
$$
where $v$ is the four-velocity. Then 
$$
q^2-m^2=2m(vl)+l^2
$$
The scalar product $(vl)\equiv E$ just has the meaning of the energy variable.
There is only one more $O(m)$ term in the propagator:
$$
2(qk)=2m(vk)\{1+O(1/m)\} \  .
$$
With this accuracy, one can write down the relevant Feynman integral as
$$
I(m,v,E)=\int \frac{d^4k}{2m(E+(vk))}S(k) \  , 
$$
where $S(k)$ is the propagator of the light particle. 
Now, applying the Borel transformation $B(E \to \epsilon)$ which converts 
$1/(E+(vk))$ into $\frac1{\epsilon}e^{-(vk)/\epsilon}$ we get
$$ M(m,v,\epsilon)\equiv B(E \to \epsilon)I(m,v,E) =
\frac1{2m\epsilon} \int  e^{-(vk)/\epsilon} S(k) d^4k.
$$
It is easy to realize that the integral above is just the light 
particle propagator in the configuration space
$$
M(m,v,\epsilon) = \frac1{2m\epsilon} \langle \varphi(0) 
\varphi({iv}/{\epsilon}) \rangle
$$
{\em i.e.,}  the Green function describing propagation in the
imaginary time, since the velocity vector $v^{\mu}$, in the rest
frame of the heavy particle, is oriented  exactly
in the  time direction.

Thus, we have demonstrated that the Borel transformed 
correlators are just the QFT generalization of the imaginary
time Green functions we studied in the quantum-mechanical examples.

\subsection{Perturbative vs nonperturbative contributions in QCD}

From the technical point of view, the strategy now is 
\\ $\bullet$  to calculate 
 the correlator $\Pi(Q^2)$ in the deep spacelike region where
one can rely on the asymptotic freedom of QCD,
\\ $\bullet$   apply the Borel transformation to the resulting expression to
get a $1/M^2$ expansion for $\Phi(M^2)$,
\\ $\bullet$   finally, using the dispersion representation for $\Phi(M^2)$, to
construct a sum rule.

The question is, what are the contributions analogous to the 
$\omega^2/\epsilon^2$ corrections? 

The standard idea about the QCD 
potential $V(r)$ is that it consists of  a Coulomb-like part 
$\sim \alpha_s(1/r^2\Lambda^2)$ calculable
perturbatively at short distances and a nonperturbative $\sim r$
long distance ``confining'' part:
$$
V(r)=V^{``Coulomb"}(r) + V^{conf}(r)
$$ 

The magnitude of the ``Coulomb'' effects is determined by the QCD running
coupling constant
$$
\alpha_s(Q^2)=\frac{4\pi}{9\log(Q^2/\Lambda^2)} + \ldots.
$$
The standard modern estimate for the value of the QCD scale $\Lambda$ is
$$
\Lambda \sim 200 \, MeV \  .
$$
 As we will see, the parameters that 
really appear in QCD sum rules have dimension of $(mass)^2$ , and 
one should compare
$$
\Lambda^2 \sim  0.04 \, {\rm GeV}^2
$$
to the typical hadronic scales like $\rho$  and nucleon masses (squared):
$$
m_{\rho}^2 \approx 0.6\, {\rm GeV}^2, \hspace{2cm} m_N^2 \approx 0.9\, {\rm GeV}^2.
$$
To get such scales from $\Lambda^2$ one needs unnaturally large numerical
factors of an order of 25 or even 100.
So, it is most unlikely that the $O(\alpha_s)$-corrections
can generate scales of an order  of $1\, 
${\rm GeV} and , hence, the ``Coulomb'' 
part  is practically irrelevant to the formation of the 
spectrum of light hadrons.
 Our problem now is to take into account the effects due to the
long-range nonperturbative part of the QCD potential.

Our study of the quantum-mechanical oscillator
(which is the simplest example of a system with confinement) has demonstrated
that the propagation amplitude in the presence
of the potential is different  from the free Green function.
Furthermore, we observed that the difference  vanishes at short
distances and that one can calculate  exact Green function
perturbatively, expanding in powers of the oscillator potential. 
In QCD, with two components of $V(r)$ one can imagine a double series,
both in $ V^{``Coulomb"}(r)$ and $ V^{conf}(r)$. 
There is no problem to arrange an expansion in
$V^{``Coulomb"}(r)$ , $-$  this will be just the standard perturbation
theory in the QCD coupling constant. On the other hand, 
the confining potential $V^{conf}(r)$ in QCD is not even known. 
Moreover, the widespread belief 
is that it is determined by some essentially
nonperturbative effects, {\em i.e.}, by those which  cannot be seen in 
perturbation theory. 

In such a situation, 
a possible way out is to proceed as follows: 
\\ $\bullet$  to construct perturbation expansion in terms of quark and gluon
propagators (this amounts to taking into account only the Coulombic
part of the potential),
\\ $\bullet$  to postulate that quark and gluon propagators are modified by the
long-range confinement part of the QCD potential; but the modification
is soft in a sense that at short distances the difference between 
exact and perturbative (free-field) propagators vanishes.

To formalize this statement, one can write the exact propagator 
${\cal D}^{exact}(x)$ as a vacuum average of a T-product of fields in the 
exact vacuum $\Omega$ (corresponding to a confining potential) 
$$
{\cal D}^{exact}(x) = \langle \Omega |\,T(\varphi(x)\varphi(0))|\Omega\rangle .
$$
According to the Wick theorem, one can write the $T-$product as the sum
$$
T(\varphi(x)\varphi(0)) = \underbrace{\varphi(x)\varphi(0)} + 
:\varphi(x)\varphi(0):
$$
of the ``pairing'' and the ``normal'' product. The ``pairing'' is just
the expectation value of the $T$- product over the perturbative vacuum
$$
\underbrace{\varphi(x)\varphi(0)} = 
\langle 0 |\,T(\varphi(x)\varphi(0))|0 \rangle .
$$
{\em i.e.}, the perturbative propagator. By this definition, the normal 
product \mbox{$:\varphi(x)\varphi(0): $} 
vanishes if averaged over the perturbative 
vacuum:
$$
\langle 0 |\,:\varphi(x)\varphi(0):|0 \rangle = 0.
$$
Thus, our assumption that 
${\cal D}^{exact}(x) \neq {\cal D}^{pert}(x)$
is  equivalent to the statement 
$$
\langle \Omega |\,:\varphi(x)\varphi(0):|\Omega\rangle \neq 0,
$$
which is the starting point to calculating  power corrections in QCD.
To simplify the notation, in what follows we will write the lhs of the 
above equation as $\langle \,\varphi(x)\varphi(0)\rangle $.

\subsection{Condensates} 

In the oscillator case the analog of 
$\langle \,\varphi(x)\varphi(0)\rangle $ is the 
difference between the exact Green function 
$$
G^{osc}(\,0,\,0\,|\,0,\, {\tau}/i)= 
\frac{m\omega}{2\pi\sinh(\omega\tau)}
$$ 
and the  free Green function 
$$
G^{free}(\,0,\,0\,|\,0,\,{\tau}/i )=\frac{m}{2\pi\tau} 
$$
where $\tau \equiv 1/\epsilon $ is just the imaginary time variable.
Note, that both oscillator and free Green function are singular
for $\tau \to 0$, but the difference 
\ba
G^{osc}(\,0,\,0\,|\,0,\,{\tau}/{i})-
G^{free}(\,0,\,0\,|\,0,\,{\tau}/{i})  \nonumber \\
=  \frac{m}{2\pi} \left [-\frac1{6}\, \omega^2\tau +\frac7{360}\, 
\omega^4\tau^3
-\frac{31}{15120}\, \omega^6\tau^5 + \ldots \right ] 
\ea
is regular at that point, and
one can even expand the difference into
 the Taylor series in $(\omega\tau)^2$.
The expansion  is possible essentially because the confining potential 
is regular (and even vanishing) at $\tau =0$. 

In the QCD  sum rule approach it is also assumed that the confinement
effects are sufficiently soft to allow for the Taylor expansion of 
$\langle \,\varphi(x)\varphi(0)\rangle $
at $x=0$:
\ba
\langle \,\varphi(0)\varphi(x)\rangle   & = & \sum_{n=0}^{\infty}
\frac{x^{\mu_1} \ldots x^{\mu_n}}{n!} 
\langle \,\varphi(0)\partial_{\mu_1} \ldots \partial_{\mu_n} \varphi(0)\rangle 
\nonumber \\
& = & \langle \,\varphi \varphi\rangle 
+x^{\mu} \langle \,\varphi \partial_{\mu} \varphi\rangle 
+\frac{x^{\mu_1}x^{\mu_2}}{2}
\langle \,\varphi \partial_{\mu_1}\partial_{\mu_2}  \varphi\rangle + \ldots
\ea
This is, in  fact, the expansion of the nonlocal object 
$\langle \,\varphi(0)\varphi(x)\rangle $ 
over the vacuum matrix elements of the local composite operators. 
Not all operators really contribute to the expansion above.
Using the Lorentz covariance, it is trivial to find that
\ba
\langle \,\varphi\partial_{\mu} \varphi\rangle & = & 0, \nonumber \\
\langle \,\varphi \partial_{\mu_1} \partial_{\mu_2} \varphi \rangle & = &
\frac1{4} g_{\mu_1 \mu_2} \langle \,\varphi\partial^2 \varphi\rangle ,
\ea
{etc.}, so that finally one arrives at the expansion in $x^2$: 
\begin{equation}
\langle \,\varphi(0)\varphi(x)\rangle = 
\sum_{n=0}^{\infty} \left (\frac{x^2}{4}\right )^n \frac1{n!(n+1)!}\, 
\langle \varphi(\partial^2)^n \varphi\rangle \, .
\end{equation}
Thus, the modification of the propagator by the  
nonperturbative effects is now parametrized by the matrix elements of 
the composite operators like
$\langle \varphi(\partial^2)^n \varphi\rangle$. 
The examples in QCD are 
\\ $\bullet$ {$\langle \bar q q \rangle$ referred to as the quark condensate,}  
\\ $\bullet$ {$\langle \bar q D^2 q \rangle$, characterizing the average virtuality
of the vacuum quarks,} 
\\ $\bullet$ {the gluon condensate
$\langle G_{\mu\nu}^a G_{\mu\nu}^a \rangle$ ,
 {etc.} }
Here $D_{\mu} \equiv \partial_{\mu} - igA_{\mu}$ is the 
covariant derivative and $G_{\mu\nu} = (i/g)[D_{\mu} , D_{\nu}]$ 
is the gluonic field strength. Note, that only gauge invariant composite
operators should appear in QCD, {\em i.e.} each $\partial_{\mu}$ must
be accompanied by the relevant $A_{\mu}$. How this happens, will be
discussed later.

\subsection{Operator product expansion}

It is instructive to analyze the above expansion in the 
momentum representation
\be
\langle \,\varphi(0)\varphi(x)\rangle \equiv
\int  e^{ikx} {\cal D}(k)\, d^4 k=
\sum_{n=0}^{\infty}  \left (\frac{x^2}{4} \right )^n 
\frac1{n!(n+1)!} \int   (-k^2)^n {\cal D}(k)\,d^4 k .
\end{equation}
Note, that existence of  $\langle \varphi(0) \varphi(0) \rangle$ implies
that ${\cal D}(k)$, the nonperturbative part of the propagator,
vanishes for large momenta faster than $k^{-5}$, while existence of 
 $\langle \varphi(0) \partial^2 \varphi(0) \rangle$ implies
that ${\cal D}(k)< k^{-7}$, {etc.} All  matrix elements 
$\langle \varphi(0) (\partial^2)^n \varphi(0) \rangle$ 
exist only if, for 
large momenta, the function 
${\cal D}(k)$ vanishes  faster than any power of $1/k^2$, say, like an 
exponential.
Thus, the function ${\cal D}(k)$ is concentrated in the region
of small momenta. This corresponds to the basic assumption
that at large momenta one should enjoy 
the asymptotic freedom.

\begin{figure}[t]
\mbox{
   \epsfxsize=12cm
 \epsfysize=3cm
  \epsffile{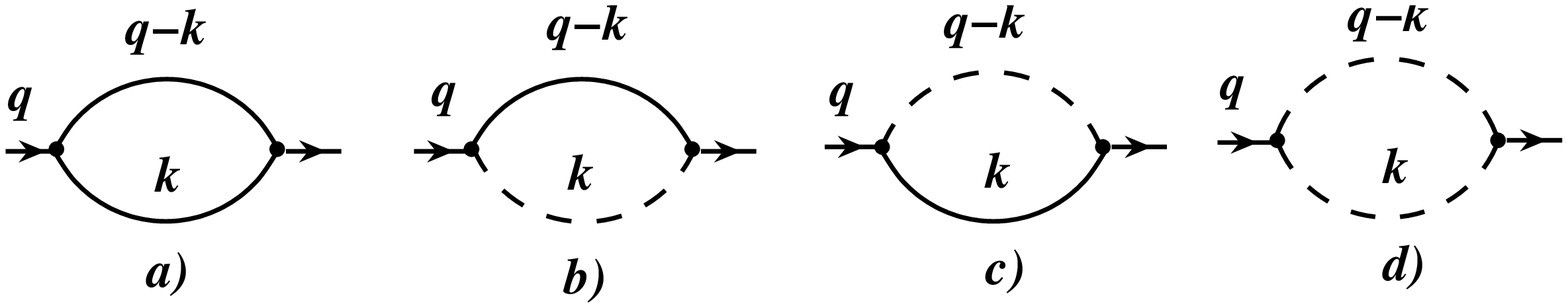}  }
{\caption{\label{fig:scalope} 
Operator product expansion for scalar one-loop diagram.}}
\end{figure}

In other words, the exact propagator is represented as a sum of the
perturbative $O(1/k^2)$ part and the nonperturbative (``condensate'')
part that vanishes fast when the momentum $k$ increases.
Using such a decomposition, we obtain a modified diagram technique
for the correlator functions.
Consider, {\em e.g.,} the lowest order diagram corresponding to the 
product of two propagators. After the decomposition it produces four
diagrams (see Fig.\ref{fig:scalope}), 
where the dashed lines correspond to the nonperturbative part of the
propagators.
 
The first term (Fig.\ref{fig:scalope}a) 
is just the ordinary purely perturbative diagram 
having $\sim \log(q^2)$ behavior for large $q^2$:
$$
\int \frac{d^4 k}{k^2 (q-k)^2} \sim \log(q^2).
$$
Two other diagrams (Figs.\ref{fig:scalope}b,c), 
as we will see below, behave like $1/q^2$.
This  behavior
is completely determined by that of the perturbative propagator:
the whole momentum flows through the perturbative line where its
propagation is not suppressed. 
In the last diagram (Fig.\ref{fig:scalope}d), all possible
momentum flows are strongly suppressed, {\em e.g.,} if ${\cal D}(k)$, 
the nonperturbative part of the propagator, 
vanishes exponentially at large $k$, so does this contribution also.

Now, let us analyze the contribution of the second diagram:
\be
I^{\langle \varphi \varphi \rangle}(q^2) =
\int {\cal D}(k^2)\frac{d^4 k}{(q-k)^2} .
\end{equation}
As we discussed earlier, the function ${\cal D}(k^2)$ is concentrated 
in the region of small momenta. So it makes sense to expand the
perturbative propagator $1/(q-k)^2$ in $k$:
\be
\frac1{(q-k)^2} = \frac1{q^2} \sum_{n=0}^{\infty} 
\frac{ 2^n \{qk\}^{(n)}}{(q^2)^n}\,  \theta(k^2<q^2) +
\frac1{k^2} \sum_{n=0}^{\infty} 
\frac{ 2^n \{qk\}^{(n)}}{(k^2)^n} \, \theta(q^2<k^2)
 \end{equation}
where 
$$
\{qk\}^{(n)} \equiv \{q_{\mu_1} \ldots q_{\mu_n}\}\{k^{\mu_1} \ldots k^{\mu_n}\}
$$ 
is the product of two symmetric 
$$
O^{\ldots \mu_i \ldots \mu_j \ldots} = O^{\ldots \mu_j \ldots \mu_i \ldots}
$$
and traceless 
$$
O^{\ldots \mu_i \ldots \mu_j \ldots}g_{\mu_i\mu_j}=0
$$ 
tensors formed from the vectors $q$ and $k$, respectively. 
The symmetric-traceless tensors have a very useful property:
$$
\int  F(k^2) \{k^{\mu_1} \ldots k^{\mu_n}\} \, d^4 k= 0
$$
for $n \neq 0 $, because it is impossible to construct a nontrivial
symmetric-traceless 
combination using only the metric tensors $g_{\mu_i\mu_j}$ . To do this,
one should have at least one vector, {\em e.g.,} 
$$
\{p_{\mu_1} p_{\mu_2}\}= p_{\mu_1} p_{\mu_2} - \frac1{4}\, p^2 \,g_{\mu_1 \mu_2}.
$$
Thus, in our case
\ba
I^{\langle \varphi \varphi \rangle}(p^2) & \equiv &
\int {\cal D}(k^2)\frac{d^4 k}{(q-k)^2}  \nonumber \\ 
& = & \frac1{q^2} \int {\cal D}(k^2) \theta(k^2<q^2)  d^4 k+
\int {\cal D}(k^2) \frac{d^4 k}{k^2} \theta(q^2<k^2) \nonumber \\
& = & \frac1{q^2}\langle \varphi \varphi \rangle + 
``quickly \ vanishing \ contribution",
\ea
where the ``quickly vanishing'' term is given 
by the integral over large momenta:
$$
``quickly  \ vanishing \ contribution" = 
\int \left (\frac1{k^2}-\frac1{q^2}\right ) \, 
{\cal D}(k^2) \, \theta(k^2>q^2)\, d^4 k 
$$
where the function ${\cal D}(k^2)$ is strongly suppressed.

Thus, it is very easy to calculate this contribution: the whole
external momentum flows through the perturbative line, so one should 
just multiply 
the perturbative propagator $1/q^2$  by the condensate  
$\langle \varphi \varphi \rangle$ factor. One can wonder, however,
why the higher condensates (containing  the $\partial^2$) did not
appear? To answer this question, it is useful to analyze this 
contribution in the configuration representation. The original diagram
decomposition now looks like 
\ba
{\cal D}(x,0){\cal D}(0,x) & = & {\cal D}^{pert}(x,0){\cal D}^{pert}(0,x)+
{\cal D}^{pert}(x,0)\langle \,\varphi(0)\varphi(x)\rangle \nonumber \\ 
& + &{\cal D}^{pert}(0,x)\langle \,\varphi(x)\varphi(0)\rangle
+\langle \,\varphi(0)\varphi(x)\rangle \langle \,\varphi(x)\varphi(0)\rangle,
\ea
and its second term can be written (modulo trivial factors like $2\pi$)  as
$$
\langle \,\varphi(0)\varphi(x)\rangle /x^2 .
$$ 
Now, using the $x^2$-expansion for the $\langle \,\varphi(0)\varphi(x)\rangle$-
factor, we obtain
\be
\frac1{x^2}\langle \,\varphi(0)\varphi(x)\rangle = 
\frac1{x^2}\langle \,\varphi \varphi \rangle +
\frac1{x^2} \frac{x^2}{8}\langle \,\varphi(0)\varphi(x)\rangle + 
O(x^2) + \ldots \ .
\end{equation}
In the momentum representation, the first term just corresponds to the
$\langle \,\varphi \varphi \rangle/q^2$ contribution, while the
second is proportional to $\delta^4(q)$ and can be discarded since 
the external momentum $q$ is assumed to be large. Higher terms are
proportional to the derivatives of  $\delta^4(q)$, and also can be ignored
for large $q$. It is easy to realize, that the sum of all these
terms just produces the ``quickly vanishing contribution'' discussed 
above. Only the terms   singular for $x^2=0$ (in the configuration 
representation) can produce a power-like behavior
in the momentum representation.

The expansion we discussed above is a simplified illustration of the
operator product expansion\cite{wilson} 
\be
T(j(x)j(0)) = \sum_{n=0} C_n(x^2) O_n(0).
\end{equation}
It represents the $T$-product of two currents $j(x), j(0)$ as 
a sum of the local (composite) operators of increasing dimension (in 
units of mass). The coefficient functions $C_n(x^2)$ characterizing
the contribution of a particular operator are ordered in their singularity
at $x^2=0$. From a simple dimensional 
analysis it follows that there should be a  simple relation
between singularity  of the coefficient function and the dimension
of the relevant composite operator; {\em e.g.,} if
$$
C_n(x^2)\sim \frac1{(x^2)^{N_0-n}}
$$
then 
$$
O_n \sim (\mu^2)^{d_0+n},
$$
so that after the Fourier transform one arrives at an expansion in
powers of $\mu^2/q^2$. There are, of course, more 
complicated contributions into the OPE, 
corresponding to more complicated original diagrams.

\begin{figure}[t]
\mbox{
   \epsfxsize=12cm
 \epsfysize=3cm
  \epsffile{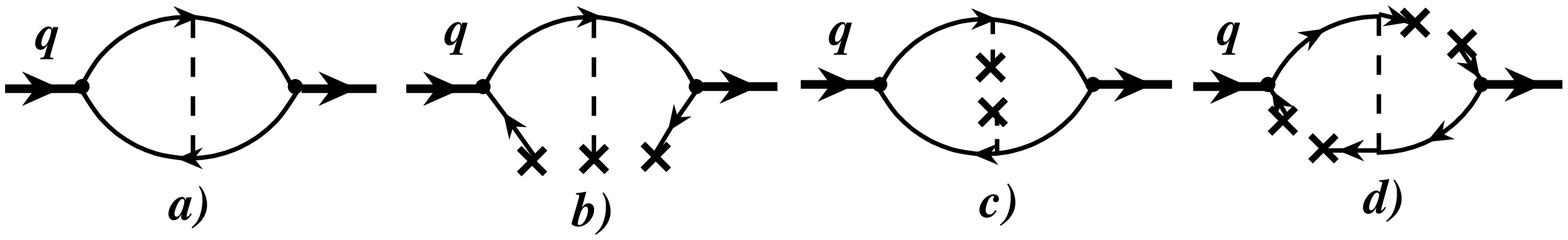}  }
{\caption{\label{fig:2lqcdope} 
Operator product expansion for QCD two-loop diagram.}}
\end{figure}

Similar decomposition of propagators into perturbative and 
nonperturbative parts can be also performed in the QCD case.
At this level of discussion, 
the only difference is that in QCD one has to
deal with two types of fields: quarks $q, \bar q$ and gluons $A$.
Consider  
the decomposition of a two-loop diagram (see Fig.\ref{fig:2lqcdope}).
\\ $\bullet$  The first diagram (Fig.\ref{fig:2lqcdope}a)
corresponds to the purely perturbative
contribution (unity operator in the OPE language), 
\\ $\bullet$  the second diagram (Fig.\ref{fig:2lqcdope}b)
(in fact there are two such diagrams) corresponds to the
$\langle \bar q A q \rangle$-type contribution,
\\ $\bullet$  the third (Fig.\ref{fig:2lqcdope}c) is evidently
proportional to the $\langle AA \rangle$ vacuum average,
accompanied by the one-loop
coefficient function, 
\\ $\bullet$  the fourth (Fig.\ref{fig:2lqcdope}d) 
diagram (there is also another such diagram) brings in 
the four-quark
vacuum average $\langle \bar qq \bar qq \rangle$.

\section{ Operator Product Expansion: Gauge Invariance and 
Related Problems}

 Looking at the operators discussed
in the previous section, one can notice that some of them,
{\em e.g.,} $\langle \bar q \partial_{\mu} q\rangle$,
$\langle \bar q A_{ \mu} q\rangle$, 
$\langle A_{ \mu}A_{ \nu} \rangle$, {etc.}, 
do not satisfy a very important requirement:  they are not gauge 
invariant. And the product of gauge invariant operators in
a gauge theory should have an expansion in  gauge
invariant operators like  
$\langle \bar q \hat D_{ \mu} q\rangle$ and
$\langle G^a_{ \mu \nu}G^{a \, \mu \nu} \rangle$.  

\subsection{ Notations}

Let us introduce now the necessary  notations. To begin with, 
$$
\hat D_{\mu} = \partial_{\mu} - ig \hat A_{\mu}
$$ 
is  the covariant derivative in the quark (fundamental) representation
of the $SU(3)_c$ gauge group:
$$
 \hat A_{\mu} =  A_{\mu}^a \tau^a,
$$
with matrices $(\tau^a)_{AB}$ simply related to the Gell-Mann $(3\times3)$
$\lambda$-matrices
$$ 
\tau^a \equiv \lambda^a/2.
$$
The indices $A,B$ correspond to 3 possible quark colors:$A,B=1,2,3$. 
The gluons have 8 colors described by the $a$-indices: $a=1, \ldots 8$.
The covariant derivative 
$$ \tilde D_{\mu}=\partial_{\mu}-ig \tilde A_{\mu}
$$
acting on the gluonic field contains the $(8\times8)$ $\sigma$-matrices
$$
 \tilde A_{\mu}= A_{\mu}^a \sigma^a
$$
characteristic for the gluonic (adjoint) representation of the
$SU(3)$ color group:
$$
(\sigma^a)_{bc} = -if_{abc},
$$
where $f_{abc}$ are the structure constants of the group:
$$
[\tau^a, \tau^b] = i f^{abc} \tau^c.
$$
The gluonic field strength tensor
$$G^a_{ \mu \nu}=\partial_{\mu}A_{\nu}-\partial_{\nu}A_{\mu}+
gf^{abc}A^b_{\mu}A^c_{\nu}
$$
can be related to the commutator of the covariant derivatives:
\ba
\hat G_{ \mu \nu} = \frac1{ig}[\hat D_{\mu}, \hat D_{\nu}] \\
\tilde G_{ \mu \nu} = \frac1{ig}[\tilde D_{\mu}, \tilde D_{\nu}].
\ea

\subsection{External field method}

\begin{figure}[t]
\mbox{
   \epsfxsize=12cm
 \epsfysize=3cm
  \epsffile{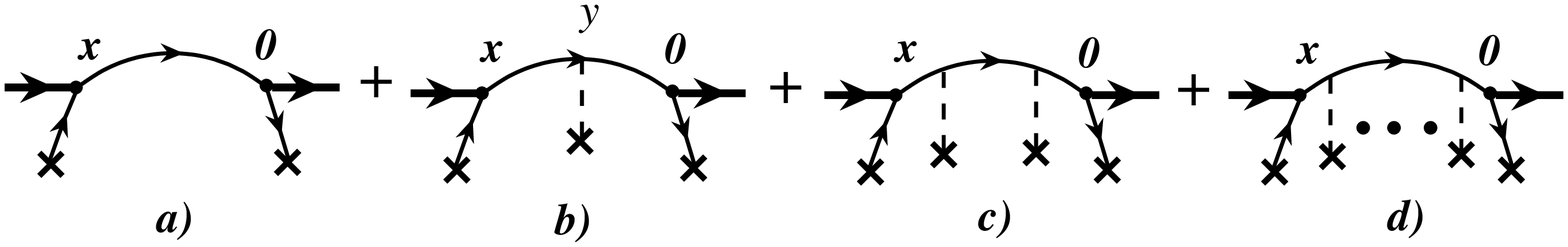}  }
{\caption{\label{fig:qcdgl} 
Summation over gluons inserted into hard quark propagator.}}
\end{figure}

A rather innocent-sounding statement 
that only  gauge invariant operators should appear 
in the operator product expansion implies that there are numerous 
nontrivial correlations between the coefficient functions 
of non-gauge-invariant operators originating from 
completely different diagrams. For example, the operator
$\langle \bar q \partial_{\mu} q\rangle$ appears when one expands
the nonlocal operator 
$\langle \bar q(0) q(x) \rangle$
in the simplest combination  (see Fig.\ref{fig:qcdgl}a)
$$
\langle \bar q(0) S(-x) q(x) \rangle,
$$
while the operator
$\langle \bar q \hat A_{ \mu} q\rangle$ necessary to form a gauge-invariant
sum, results from a more complicated combination(see Fig.\ref{fig:qcdgl}b)
$$
\int  \langle \bar q(0) S(-y)\gamma^{\mu} (ig) \hat A_{\mu}(y) 
S(y-x) q(x) \rangle \, d^4y .
$$
In both cases, the perturbative propagators (in configuration space) 
are written explicitly,
while the nonperturbative parts, in accordance with
our previous discussions, are denoted by vacuum 
averages of the relevant normal
products of fields. By a straightforward 
 integration one can establish that
$$
\int S(-y)\gamma^{\mu} S(y-x)\, d^4y = -x^{\mu} S(-x) \, ,
$$
and, hence, the coefficient functions of the two operators are
just those necessary to form the gauge invariant combination.

However, this is definitely not the method one would like to use
in more complicated situations. In fact, it is much easier
to analyze the sum of all the combinations differing only by the number
of the gluons going into vacuum (see Fig.\ref{fig:qcdgl}).

One can write it as\cite{ER3,nikrady}
$ \langle \bar q(0) S(0,x;A) q(x) \rangle  \, , 
$
where $S(0,x;A)$ has the meaning of the perturbative quark propagator
in the (external) vacuum gluonic field $A$. It satisfies the Dirac equation
\be
\left [i\gamma^{\mu}(\frac{\partial}{\partial x^{\mu}} - ig\hat A_{\mu}(x))
-m \right ] S(x,y;A) = - \delta^4(x-y).
\end{equation}
Solving it by using perturbation theory in $A$, one gets back the 
original diagrammatic expansion. 

The function $S(x,y;A)$ is not gauge invariant. From the good old 
electrodynamics 
it is known that propagation of a charged particle in the external field
results in a phase factor  given by the exponential of a line integral.
In the QCD case it looks like
$$
\hat P(x,y;A;{\cal C})=P\exp(ig\int_{\cal C} \hat A_{\mu}(z) \,dz^{\mu}),
$$
where $P$ means that the color matrices should be ordered along the
path ${\cal C}$ connecting the points $x$ and $y$. 
Furthermore, it is known that the combination
$$
\langle \bar q(y) \hat P(y,x;A;{\cal C}) q(x) \rangle
$$
is gauge invariant (though path-dependent). 
This gives the idea how one can construct the
operator product expansion in
an explicitly gauge invariant way.
Let us try the ansatz
\be
S(x,y;A) = \hat E(x,y;A) {\cal S}(x,y;A)
\end{equation}
where $\hat E(x,y;A)$ is the $P$-exponential for the straight-line path
connecting the points $x$ and $y$:
$$
\hat E(x,y;A)= P\exp \left [ig(x^{\mu}-y^{\mu})
\int_0^1 \hat A_{\mu}(z) \,dt \right ]_{z=y+t(x-y)}.
$$
 
It is straightforward to derive that the original Dirac equation is 
satisfied only if the function ${\cal S}(x,y;A)$ is a solution
to the modified Dirac equation
\be
\left [i\gamma^{\mu}\left (\frac{\partial}
{\partial x^{\mu}} - ig\hat{\cal A}_{\mu}(x;y) \right )
-m \right ]\, {\cal S}(x,y;A) = - \delta^4(x-y) 
\end{equation}
that differs from the original one only by the change
\be
A^a_{\mu}(x) \to {\cal A}^a_{\mu}(x;y)=
(x^{\nu}-y^{\nu})\int_0^1 G_{\nu\mu}^b(z)\,\tilde E^{ba}(z,y) \,tdt.
\end{equation}
Here, $z=y+t(x-y)$ and $\tilde E$ is a straight-line-ordered exponential with the
$A$-fields multiplied by the $\sigma$-matrices  of the gluonic
representation. It appears after one commutes the $\hat E$-exponentials
through a $\tau$-matrix:
\be
(\tau^a)_{AB}\hat E_{BC}(z,y)=\hat E_{AB}(z,y)(\tau^b)_{BC}\tilde E^{ab}(z,y).
\end{equation}
The commutation rule is based on the well-known formula
\be
e^ABe^{-A}=B+[A,B]+\frac1{2!}[A,[A,B]]+\ldots
\end{equation}
and the relation
\be
[\tau_a,\tau_b]=-(\sigma_a)_{bc}\tau_c.
\end{equation}

Surprizingly enough, the presence of the $\tilde E$-factor only
simplifies the situation, because the derivatives in the Taylor
expansion of the $G\tilde E$-combination are just the covariant ones: 
\be
G_{\nu\mu}^b(z)\,\tilde E^{ba}(z,y)=
\sum_{n=0}^{\infty} \frac1{n!}\, G^a_{\nu\mu;\mu_1 \ldots \mu_n}(y)
(z-y)^{\mu_1}\ldots(z-y)^{\mu_n}.
\end{equation}
As a result, the new field ${\cal A}^a_{\mu}(x;y)$  can be expressed
in terms of the gluon field stress tensor $G_{\mu\nu}$ and its
covariant derivatives (taken, of course, in the gluonic representation):
\be
{\cal A}^a_{\mu}(x;y)=
\sum_{n=0}^{\infty} \frac1{(n+2)n!} (x-y)^{\nu}
(x-y)^{\mu_1}\ldots(x-y)^{\mu_n}\,G^a_{\nu\mu;\mu_1 \ldots \mu_n}(y).
\end{equation}

This means that the function ${\cal S}(x,y;A)$ depends on the gluonic
field only through the gluon field stress tensor $G_{\mu\nu}$ and its
covariant derivatives. Solving the equation for ${\cal S}(x,y;A)$ 
perturbatively, one obtains the $g{\cal A}$-expansion that has the same 
structure as the original $gA$-expansion. A very important difference
is that the resulting operators have an explicitly gauge invariant form.
For example, taking the lowest (zero) order term one obtains the contribution
\be
\langle \bar q(y) S(y-x)\hat E(y,x;A) q(x) \rangle 
\end{equation}
containing the gauge-invariant operator
\be
\langle \bar q(y) \hat E(y,x;A) q(x) \rangle =
\sum_{n=0}^{\infty} \frac1{n!} (x-y)_{\mu_1}\ldots (x-y)_{\mu_n}
\langle \bar q \hat D^{\mu_1}\ldots \hat D^{\mu_n} q \rangle|_y .
\end{equation}

We used the direct straight-line path from $x$
to $y$ in our ansatz. In general, it is more
convenient to use the path formed by two straight lines: from $x$ to
some fixed point $z_0$ (usually taken at the origin, $z_0=0$) and then
from $z_0$ to $y$. The only change then will be that quark and gluon 
fields $q(\xi)$, $A(\xi)$ will be finally accompanied by  the $E(\xi,z_0;A)$ 
exponentials, so that a covariant expansion will be obtained if
one expands at the fixed point $z_0$. Using translation invariance,
one can always substitute  matrix elements 
$\langle \bar q(z_0)\ldots G(z_0)\ldots q(z_0)\rangle$, resulting from such
a procedure, by $\langle \bar q(0)\ldots G(0)\ldots q(0)\rangle$.
Less trivial is the disappearance of 
the $z_0$-dependence from the coefficient functions. One can enjoy it
only if the calculations had no errors, and after summation over all diagrams
(of the modified expansion) producing the same operator. The latter
observation suggests that $z_0$ works as a gauge parameter.
Indeed, it is easy to notice that the relevant $P$-exponentials
\be
E(x,z_0;a)=P\exp[ig(x^{\mu}-z_0^{\mu})\int_0^1 A_{\mu}(z) \,dt]
|_{z=z_0+t(x-z_0)}
\end{equation}
disappear (are equal to 1) if one imposes the gauge 
condition\cite{fock,schwinger,ds,shifman,nikrady}
\be
(x^{\mu}-z_0^{\mu})A_{\mu}(x)=0.
\end{equation}
Furthermore, the ``calligraphic'' field ${\cal A}$ in this 
(Fock-Schwinger or fixed point) gauge coincides 
with the ordinary one $A={\cal A}(\{G\})$, {\em i.e.},
in this gauge not only $G_{\mu\nu}$ can be expressed in terms
of $A_{\mu}$, but also the vector-potential $A_{\mu}$ can
be represented as a (nonlocal) functional of the field strength
tensor $G_{\mu\nu}$. This means that the Fock-Schwinger gauge
is a physical gauge. Another example is the axial gauge\cite{ikr} 
\be
n_{\mu}A^{\mu}(x)=0
\end{equation}
with $n_{\mu}$ being a fixed vector. In this gauge one can also
express\cite{ikr}  $A_{\mu}$ in terms of $G_{\mu\nu}$:
\be
A_{\mu}(x)= n^{\nu} \int_0^{\infty} G_{\nu\mu}(x+tn)\,dt.
\end{equation}

Imposing the Fock-Schwinger gauge condition, one can forget about
the exponentials. One should remember, however, that the ``ordinary''
derivatives in this gauge should be treated as the covariant ones,
{\em e.g.,} they do not commute, {etc.} 
To illustrate the typical
tricks and conventions, let us consider the simplest 
gauge invariant nonlocal condensate
\ba
\langle \bar q(0) \hat E(0,x;A) q(x) \rangle|_{(xA)=0} =
\langle \bar q(0) q(x) \rangle  \nonumber \\
= \sum_{n=0}^{\infty} \frac1{n!} x_{\mu_1}\ldots x_{\mu_n}
\langle \bar q \hat D^{\mu_1}\ldots \hat D^{\mu_n} q \rangle 
= \langle \bar qq \rangle +
\frac{x^2}{8} \langle \bar q \hat D^2 q \rangle +\ldots .
\ea
The ratio 
\be
\lambda^2 \equiv \frac{\langle \bar q \hat D^2 q \rangle}
{\langle \bar qq \rangle }  
\end{equation}
can be interpreted as the average virtuality of the vacuum quarks. 
It is possible, however, to represent the 
$\langle \bar q \hat D^2 q \rangle$ 
operator in a different form:
\ba
\langle \bar q \hat D^2 q \rangle \equiv 
\langle \bar q \hat D^{\mu} \hat D_{\mu} q \rangle
=\langle \bar q \hat D^{\mu} \hat D^{\nu} q \rangle g_{\mu\nu}=
\langle \bar q \,\hat \slash \hspace{-2.5mm}D 
\,\hat \slash \hspace{-2.5mm}D  q \rangle -
\langle \bar q \hat D^{\mu}\hat D^{\nu}\sigma_{\mu\nu} q \rangle \nonumber \\
= - m_q^2 \langle \bar q  q \rangle -
\frac1{2}\langle \bar q\, [\hat D^{\mu},\hat D^{\nu}]\,\sigma_{\mu\nu} q \rangle
= - m_q^2 \langle \bar q  q \rangle  +
\frac1{2}\langle \bar q ig\hat G^{\mu\nu} \sigma_{\mu\nu} q \rangle
\ea
where we have used the identity
$$
g_{\mu\nu}= \gamma_{\mu}\gamma_{\nu} - 
\frac{\gamma_{\mu}\gamma_{\nu}-\gamma_{\nu}\gamma_{\mu}}{2} = 
\gamma_{\mu}\gamma_{\nu} -\sigma_{\mu\nu},
$$
the equation of motion
\be
\hat \slash \hspace{-2.5mm}D q(x) =-im_q q(x)
\end{equation}
and the definition of the field strength tensor
\be
[\hat D_{\mu}, \hat D_{\nu}]=-ig \hat G_{ \mu \nu} .
\end{equation}
Thus, the ``average virtuality'' of the vacuum quarks 
$\langle \bar q \hat D^2 q \rangle$
is directly related to the ``average vacuum gluonic field strength'' 
$\langle \bar q  ig \hat G^{\mu\nu} \sigma_{\mu\nu} q \rangle$ .
In many papers one can find the notation
\be
\langle \bar q  ig\hat G^{\mu\nu} \sigma_{\mu\nu} q \rangle \equiv
m_0^2 \langle \bar q q \rangle.
\end{equation}
Using it, one can write the relation 
\be
\lambda^2 = \frac{m_0^2}{2} -m_q^2.
\end{equation}
For  light quarks, the $m_q^2$ term can be neglected.

\subsection{ Equations of motion and the current conservation}

The equation
of motion $\hat \slash \hspace{-2.5mm}D q(x) =-im_q q(x)$ 
brings in correlations between 
matrix elements of different operators. These correlations, in fact,
are very important to preserve the symmetry properties required
by the structure of the currents $j(x)$ in the correlators. For example,
one might expect that a correlator of the electromagnetic vector currents
\be
\Pi_{\mu\nu}(p) = \int \langle T\, j_{\mu}(0)j_{\nu}(x)\rangle e^{ipx}\,d^4x
\end{equation}
should be transverse:
\be
\Pi_{\mu\nu}(p) = (p_{\mu}p_{\nu}-g_{\mu\nu}p^2)\,\Pi(p^2)
\end{equation}
due to the current conservation condition $\partial^{\mu}j_{\mu}=0$.
Indeed, for the perturbative one-loop contribution, using the explicit
form of the  propagator in the configuration space
\be
S(x) \propto -2i\frac{\slash \hspace{-2mm}x}
{x^2} \left [\frac1{x^2}+\frac{m^2}{4} \right ]+\frac{m}{x^2}+O(m^3)
\end{equation}
one finds that the integrand is
\be
\frac1{(x^2)^4}\left [g_{\mu\nu}x^2-2x_{\mu}x_{\nu}\right ] 
- \frac{m^2}{(x^2)^3}
\left [g_{\mu\nu}x^2-4x_{\mu}x_{\nu}\right ]+\ldots  \  .
\end{equation}
It is easy to see that both terms satisfy the transversality condition
$$
\frac{\partial}{\partial x^{\mu}}\, \{\ldots\}=0,
$$  
and this means that
$$\Pi^{pert}_{\mu\nu}(p) = (p_{\mu}p_{\nu}-g_{\mu\nu}p^2)\Pi^{pert}(p^2)$$
at least up to the third order in $m$. One can, in fact, perform 
the calculations  in the momentum space directly, to see that this property
of $\Pi^{pert}_{\mu\nu}(p)$ is valid to all orders in $m$. 

Now, let us consider the simplest nonperturbative contribution.
In the momentum representation it looks like
\be
\langle \bar q \gamma_{\mu}\frac{\slash \hspace{-2mm}p +m}{p^2-m^2}
\gamma_{\nu} q \rangle \sim \frac{mg_{\mu\nu}}{p^2}\langle \bar q q \rangle
\ . 
\end{equation}
However, it seems that there is no way to get the $p_{\mu}p_{\nu}$ term,
since there is only one $\slash \hspace{-2mm}p$ term in the numerator.
To solve this puzzle, let us consider this contribution in the
configuration representation. Up to $O(m^2)$-terms one has
\ba
\langle \bar q(0)\gamma_{\mu} S(x)\gamma_{\nu} q(x) \rangle\propto
\langle \bar q(0)\gamma_{\mu} 
\left (-2i\frac{\slash \hspace{-2mm}x}{(x^2)^2} + \frac{m}{x^2} \right )
\gamma_{\nu} q(x) \rangle +O(m^2) \nonumber \\ 
=-\frac{2i}{(x^2)^2}
\langle \bar q(0)\gamma_{\mu}\slash \hspace{-2mm}x  \gamma_{\nu} q(x) \rangle 
+  \frac{m}{x^2} \langle \bar q(0)\gamma_{\mu} \gamma_{\nu} q(x) \rangle
+O(m^2).
\ea
Note, that vacuum averages of local operators 
with an odd number of indices vanish,
and only those with an even number survive. Incorporating the Lorentz 
invariance, we obtain
\ba
\langle \bar q(0)\gamma_{\mu} \gamma_{\nu} q(0)\rangle &=&
g_{\mu\nu} \langle \bar q q \rangle, \nonumber \\
\langle \bar q(0)\gamma_{\mu}\gamma_{\alpha} \gamma_{\nu} q(0)\rangle &=& 0,
\nonumber \\
\langle \bar q(0)\gamma_{\mu}\gamma_{\alpha} \gamma_{\nu} D_{\beta}q(0)\rangle 
&=&(g_{\mu\alpha}g_{\nu\sigma}+g_{\nu\alpha}g_{\mu\sigma}-
g_{\mu\nu}g_{\alpha\sigma})\langle \bar q\gamma_{\sigma}D_{\beta} q\rangle  ,
\nonumber \\ 
\langle \bar q\gamma_{\sigma}D_{\beta} q\rangle 
&=&\frac{g_{\sigma\beta}}{4}\langle \, 
\bar q\,\hat \slash \hspace{-2.5mm}D q\rangle .
\ea
Finally, incorporating the equation of motion 
\begin{equation}
\langle \bar q\,\hat \slash \hspace{-2.5mm}D q\rangle=
-im\langle \bar q q\rangle
\end{equation}
one obtains the transverse result 
\be
\langle \bar q(0)\gamma_{\mu} S(x)\gamma_{\nu} q(x) \rangle \propto
\frac{m}{2(x^2)^2}\langle \bar q q\rangle [g_{\mu\nu}x^2+2x_{\mu}x_{\nu}] .
\end{equation}

It is instructive to analyze also the $\langle GG \rangle$ contribution.
First, using the exponential representation, it is easy to see that for
the closed loop the exponentials corresponding to the upper and the lower
lines cancel each other, and only the $O(G)$ terms remain. This general
result can be illustrated on the example of one-loop diagrams
with two vacuum gluons. To study,  how the cancellation proceeds 
in this case, one should  expand the exponential up to $O(A^2)$ terms:
\ba
S(x,A) = S(x) \left \{1+ig(xA)+\frac{(ig)^2}{2}(xA)^2 
+\ldots \right \}, \nonumber \\
S(-x,A)= S(-x) \left \{1-ig(xA)+\frac{(ig)^2}{2}(xA)^2 +\ldots  \right \}.
\ea
Now, combining the contributions of the three relevant diagrams, one obtains
\be
(ig)^2  \left \{\frac1{2}(xA)^2+\frac1{2}(xA)^2-(xA)^2 \right \} =0,
\end{equation}
{\em i.e.}, the $A-$dependence disappeared. To calculate the $O(G)$- terms,
it is most convenient to use the Fock-Schwinger gauge $(xA)=0$, in which
\ba
 A_{\mu}(x)=x^{\nu} \int_0^1 G_{\nu\mu}(tx)\,tdt 
= \sum_{n=0}^{\infty} \frac{x^{\nu}x^{\mu_1}\ldots x^{\mu_n}}{(n+2)n!}
G_{\nu\mu;\mu_1\ldots\mu_n}(0) \nonumber \\ 
=\frac1{2}x^{\nu} G_{\nu\mu}(0) +\ldots .
\ea
Using this representation, one can construct the Feynman rules for the
vertices corresponding to $G_{\mu\nu}$, $G_{\mu\nu;\mu_1}$, {etc.}
For example, in the momentum representation
\be
\frac{x_{\nu}}{2}G_{\nu\mu} 
\Rightarrow \frac1{2i} G_{\nu\mu} \frac{\partial}{\partial k_{\nu}},
\end{equation}
and the insertion of such a vertex into the quark propagator
gives
\be
\frac1{2i}\frac{\slash \hspace{-2mm}k}{k^2}\gamma_{\mu}
G_{\nu\mu}\frac{\partial}{\partial k_{\nu}}
\frac{\slash \hspace{-2mm}k}{k^2}
=\frac{i}{2}\,G_{\nu\mu}\frac{\slash \hspace{-2mm}k}{k^2}\gamma_{\mu}
\frac{\slash \hspace{-2mm}k}{k^2}\gamma_{\nu} 
\frac{\slash \hspace{-2mm}k}{k^2} \  .
\end{equation}
The result looks like an insertion of two $A$-fields, $A_{\nu}$ and
$A_{\mu}$, antisymmetrized with respect to the 
interchange of $\mu$ and $\nu$.
There is an important thing to remember  while 
constructing the diagrammatic technique for the $G$-type
vertices: in the momentum representation they  
contain  derivatives  acting onto 
all the propagators between the $G$-vertex and the Fock-Schwinger 
zero point. The $G$-inserted propagator looks simpler
in the configuration space:
\be
S(x,G) = - \frac{\slash \hspace{-2mm}x}{2\pi^2(x^2)^2}
-\frac{g}{16\pi^2x^2}\,x_{\alpha}\epsilon_{\alpha\beta\mu\nu}
\hat G^{\mu\nu}\gamma_{\beta}\gamma_5 + O(z_0)\langle GG \rangle +\ldots
\end{equation}
where $z_0$ is the Fock-Schwinger fixed point. If one takes $z_0=0$,
then the $O(\langle GG \rangle)$ term disappears. This means that
in the gauge $(xA)=0$ only the diagram with the $G$-insertions into
different lines survives. Its contribution in the configuration
space is easily calculated by just multiplying the two $S(x,G)$ 
propagators. Taking the trace over the $\gamma$-matrices, one 
obtains the result
\be
-\frac{2i}{3(16\pi^2)^2(x^2)^2}(g_{\mu\nu}x^2+2x_{\mu}x_{\nu})
\langle g^2 G^2 \rangle
\end{equation}
proportional to the same transverse structure as the contribution due to the
$m_q\langle \bar q q \rangle$ operator. This is because both operators
have the same dimension, equal to 4 in mass units.

The QCD sum rules for the ``standard'' mesons, like $\rho, \pi, K, etc.$,
are based on the OPE including the operators of dimension 6, at least.
One of these operators can be easily obtained from the Taylor expansion
of the nonlocal quark condensate containing a $\gamma$-matrix
\be
\langle \bar q(0) \gamma_{\sigma} q(x)\rangle =
\langle \bar q \gamma_{\sigma} \hat D_{\alpha}q \rangle x^{\alpha} +
\frac1{3!}\langle \bar q \gamma_{\sigma} 
\hat D_{\alpha_1}\hat D_{\alpha_2}\hat D_{\alpha_3}q \rangle 
x^{\alpha_1}x^{\alpha_2}x^{\alpha_3} + \ldots  \ .
\end{equation}
We recall that  the terms, containing an even number of derivatives,
vanish since the total number of indices is odd in these cases.
Incorporating the Lorentz invariance, we obtain
\ba
\langle \bar q(0) \gamma_{\sigma} q(x)\rangle &=&
\frac{x^{\sigma}}{4}\, \langle \bar q \,\hat \slash \hspace{-2.5mm}D q \rangle 
+\frac{x^{\sigma}x^2}{144}(\langle \bar q\,
\hat \slash \hspace{-2.5mm}D \hat D^{\alpha} \hat D_{\alpha} q \rangle  
\nonumber \\
& & + \langle \bar q\,\hat D^{\alpha}
\,\hat \slash \hspace{-2.5mm}D \hat D_{\alpha}\rangle 
+\langle \bar q\,\hat D^{\alpha}\hat D_{\alpha}
\,\hat \slash \hspace{-2.5mm}D  \rangle).
\ea
Next step is to use the translation invariance
$
\langle \partial^{\mu} \bar q(0) \ldots q(0) \rangle = 0 \ , 
$
the equation of motion $ \hat \slash \hspace{-2.5mm}D q = -im_q q$ and 
the commutation relation $[\hat D_{\alpha}, \hat D_{\mu}]
=ig\hat G_{\alpha\mu}$
to get 
\be
\langle \bar q(0) \gamma_{\sigma} q(x)\rangle= 
\frac{-imx^{\sigma}}{4} \biggl (\langle \bar q q \rangle +
\frac{x^2}{12} \langle \bar q \hat D^2 q \rangle \biggr )
+\frac{ig}{288}x^{\sigma}x^2 
\langle \bar q \gamma_{\mu}[\hat D_{\alpha}, \hat G^{\mu\alpha}] q \rangle.
\end{equation}
The commutator $[\hat D_{\alpha}, \hat G^{\mu\alpha}]$ is equal to the
covariant derivative of the gluonic field
\be
[\hat D_{\alpha}, \hat G^{\mu\alpha}]= 
\hat G^{\mu\alpha}\,_{;\alpha} \ .
\end{equation}
The final step is to use the gluonic equation of motion
\be
\hat G^{\mu\alpha}\,_{;\alpha}= j^{\mu}_a \tau^a 
\equiv \sum_{i=u,d,s} \langle \bar q_i \gamma^{\mu} \tau_a q_i \rangle \tau^a 
\end{equation}
analogous to the Maxwell equation.
Thus, due to the equations of motion, 
the apparently two-quark operator produces a four-quark term:
\be
\langle \bar q(0) \gamma_{\sigma} q(x)\rangle= 
\frac{-imx^{\sigma}}{4} (\langle \bar q q \rangle +
\frac{x^2}{12} \langle \bar q(0) \hat D^2 q \rangle)
+\frac{ig}{288}x^{\sigma}x^2 
\langle \bar q \gamma_{\mu} j^{\mu}_a \tau^a q \rangle.
\end{equation}

\section{ QCD Sum Rules for  $\rho$-meson} 

\subsection{ Dispersion relation}

As pointed out by SVZ (whose analysis\cite{SVZ}
we closely follow in this section) 
we  closely to determine the basic parameters
characterizing the $\rho$-meson, {\em e.g.}, its mass $m_{\rho}$ and 
its $e^+e^-$ coupling constant $g_{\rho}$,  one should consider
the correlator of two vector currents 
 \be
j_{\mu}^{(\rho)} = \frac1{2} (\bar u\gamma_{\mu}u - \bar d\gamma_{\mu}d) 
\end{equation}
with the $\rho$-meson quantum numbers:
\be
\Pi^{(\rho)}_{\mu\nu}(q) =
(q_{\mu}q_{\nu}-q^2g_{\mu\nu})\Pi^{(\rho)}(q^2) = 
i \int e^{iqx}  \langle 0 | T \,(j_{\mu}^{(\rho)}(x)\,j_{\nu}^{(\rho)}(0)
\,)|\,0\rangle\, d^4 x.
\end{equation}

 { Via} the dispersion relation
\be
\Pi^{(\rho)}(q^2=-Q^2) = \Pi^{(\rho)}(0)- 
\frac{Q^2}{12\pi^2} \int_0^{\infty} \frac{R^{I=1}(s)}{s(s+Q^2)} \, ds 
\end{equation}
this correlator is related to the total 
cross section of the $e^+e^-$-annihilation
to hadrons with the isospin $I=1$, measured in units of the 
$e^+e^- \to \mu^+ \mu^-$ cross section :
\be
R^{I=1}=\frac{\sigma(e^+e^-\to hadrons, I=1)}
{\sigma(e^+e^-\to \mu^+\mu^-)} \, .
\end{equation}
Indeed, the $j_{\mu}^{(\rho)}$ coincides with the isovector part
of the electromagnetic current and is responsible, therefore,
for the production of the hadrons with $I=1$ in the $e^+e^-$-annihilation.
The correlator $\Pi^{(\rho)}_{\mu\nu}(q)$ contains, of course,
 only the hadronic part of the relevant factors, and that explains
the division by $\sigma(e^+e^-\to \mu^+\mu^-)$ .
 
\subsection{ Operator product expansion}

\begin{figure}[t]
\mbox{
   \epsfxsize=12cm
 \epsfysize=3cm
  \epsffile{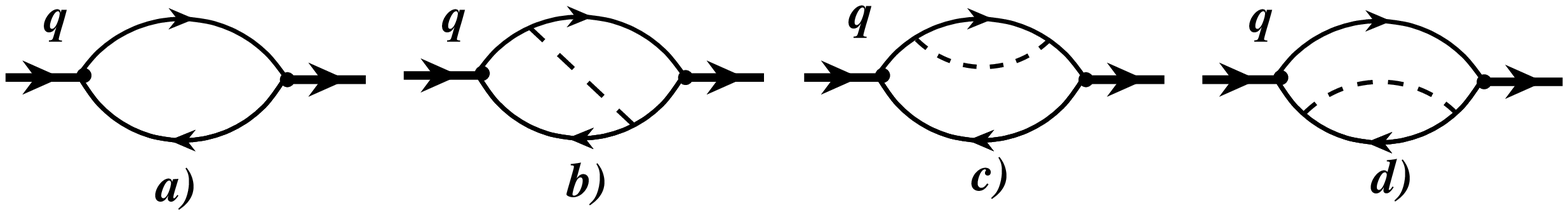}  }
{\caption{\label{fig:ropert} Perturbative contribution up to two loops.}}
\end{figure} 

\begin{figure}[t]
\mbox{
   \epsfxsize=12cm
 \epsfysize=5cm
  \epsffile{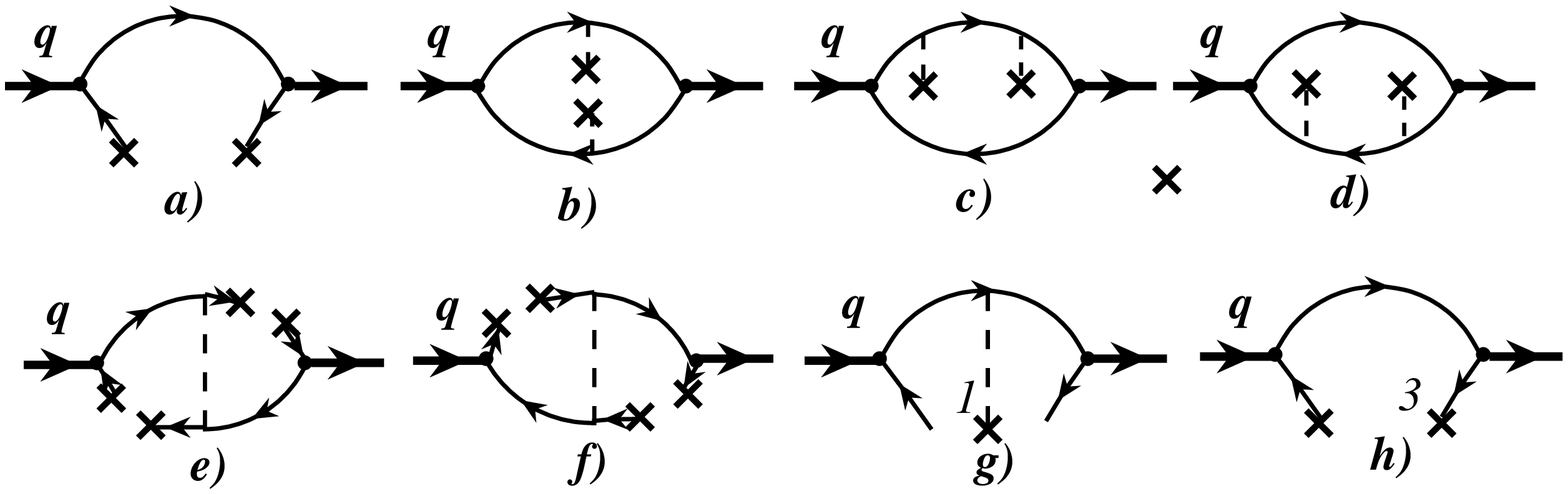}  }
{\caption{\label{fig:rocond} 
Nonperturbative contributions up to dimension six.}}
\end{figure}

On the other hand, one can construct  
the operator product expansion for  the correlator 
$\Pi^{(\rho)}_{\mu\nu}(q)$ calculating the relevant Feynman diagrams.
To begin with, one should calculate the perturbative diagrams shown
in Fig.\ref{fig:ropert}
to extract the perturbative spectral density
\be
\rho^{pert}(s)=\frac3{2}\,\theta(s)\, \left \{1+\frac{\alpha_s}{\pi}\right \}
\end{equation}
which is essentially the isovector ($I=1$ isospin projection)
quark charge  
squared $(\frac1{2})^2$ multiplied by 2 ($u$- and $d$- quarks) 
and multiplied by the number of colors (3). 
The simplest nonperturbative contribution is due to the quark
condensate (see Fig.\ref{fig:rocond}a).  
It is proportional to the current quark masses:
$$
\Pi^{\bar qq} \sim \frac{\langle 0 
|m_u\bar uu + m_d\bar dd |0\rangle}{(Q^2)^2},
 $$
with $m_u \sim 4 MeV, 
m_d \sim 7 MeV$. Hence, this contribution is rather small numerically.
The contribution 
$$\Pi^{GG}\sim g^2 \frac{\langle 0 | GG | 0 \rangle}{(Q^2)^2}$$
due to the gluon condensate is given by 
 three diagrams shown in Fig.\ref{fig:rocond}b-d.
As mentioned at the end of the previous section, 
 only the diagram \ref{fig:rocond}b
contributes in the Fock-Schwinger gauge $x_{\mu}A^{\mu}(x)=0$.
It is rather straightforward  to calculate the diagrams \ref{fig:rocond}e,f
proportional to 
$$ \frac{g^2\langle\bar qq \bar qq \rangle}{(Q^2)^3}. $$
However, the contributions of the same 
four-quark type can be obtained
also from the diagrams \ref{fig:rocond}g,h
which originally are bilinear in quark fields. 
Notice, however, the numbers 1 and  3  attached to the 
relevant lines going into the 
vacuum. 
They indicate how many  covariant derivatives are acting on the 
relevant ``vacuum'' field. Thus, the matrix elements are 
$ \langle \bar q\gamma_{\mu}\tilde D_{\nu}\hat G_{\alpha\beta} q \rangle$ 
and $\langle \bar q\gamma_{\mu}
\hat D_{\nu_1}\hat D_{\nu_2}\hat D_{\nu_3} q \rangle$.
As discussed at the end of Section 4, after vacuum averaging and 
commutations only the 
$ \langle \bar q\gamma_{\mu}\tilde D_{\nu}\hat G_{\mu\nu} q \rangle$ 
component survives and, by equations of motion it is reduced to 
$ \langle 0 |\bar q\gamma_{\mu}\tau^a j^a_{\mu} q | 0 \rangle$ 
with 
$$j^a_{\mu}= \sum_{i=u,d,s} 
\langle 0 | \bar q_i\gamma_{\mu}\tau^a q_i |0 \rangle.$$

\subsection{ Basic QCD sum rule for the $\rho$-meson channel} 

After the Borel transformation that
modifies the dispersion  integral
\be
\int \frac{\rho(s) ds}{s+Q^2} \to \int  e^{-s/M^2} \rho(s)\, \frac{ds}{M^2}
\end{equation}
and the power corrections
\be
\frac1{(Q^2)^n} \to \frac1{(n-1)!(M^2)^n}
\end{equation}
one obtains the following QCD sum rule\cite{SVZ}:
\ba
\int_0^{\infty} e^{-s/M^2} R^{I=1}(s)ds = 
\frac3{2} M^2 \biggl [1+\frac{\alpha_s(M)}{\pi}+
\frac{4\pi^2}{M^4}(\langle m_u\bar uu\rangle + \langle m_d\bar dd \rangle)+
 \nonumber \\
+\frac{\pi^2}{3M^4}\langle \frac{\alpha_s}{\pi}GG \rangle -
\frac{2\pi^3}{M^6}\langle \alpha_s(\bar u\gamma_{\alpha}\gamma_5 \tau^a u 
- \bar d\gamma_{\alpha}\gamma_5 \tau^a d)^2 \rangle -
\nonumber \\ \hspace{-2cm} 
- \frac{4\pi^3}{9 M^6} \langle \alpha_s (\bar u\gamma_{\alpha}\tau^a u 
+ \bar d\gamma_{\alpha} \tau^a d) \sum_{i=u,d,s} 
\langle 0 | \bar q_i\gamma_{\mu}\tau^a q_i |0 \rangle ].
\ea
It is the starting point for  further analysis.

\subsection{ Fixing the condensates}

 Next step is to specify the numerical values of the matrix elements 
on the r.h.s. of this sum rule (the same values will be used in all other
 sum rules). 
 \\ $\bullet$  The first matrix element is in fact fixed by the current algebra:
\be
\langle m_u\bar uu\rangle + \langle m_d\bar dd \rangle =
 - \frac1{2} f_{\pi}^2 m_{\pi}^2 = -1.7 \cdot 10^{-4} {\rm GeV}^4
\end{equation}
\\ $\bullet$  The value of the gluonic condensate was extracted from the
analysis of the QCD charmonium sum rules\cite{SVZ} 
\be
\langle \frac{\alpha_s}{\pi}GG \rangle  = 1.2 \cdot 10^{-2} {\rm GeV}^4.
\end{equation}
\\ $\bullet$  The four-quark operators are first approximated by the value 
dictated by the vacuum dominance (or factorization) hypothesis: 
\be
\langle \bar \psi_{\{A\}}\psi_{\{B\}} \bar \psi_{\{C\}} \psi_{\{D\}} \rangle = 
 \langle \bar \psi_{\{A\}}\psi_{\{B\}} \rangle 
 \langle \bar \psi_{\{C\}}\psi_{\{D\}} \rangle-
 \langle \bar \psi_{\{A\}}\psi_{\{D\}} \rangle 
 \langle \bar \psi_{\{C\}}\psi_{\{B\}} \rangle
\end{equation} 
where the subscripts $A,B,C,D$ include spin, color and flavour. Using
the orthogonality relation
\be
  \langle \bar \psi_{\{A\}}\psi_{\{B\}} \rangle = \frac{\delta_{AB}}{N}
 \langle \bar \psi\psi \rangle 
\end{equation}
one can reduce then any four quark operator to the 
$\langle \bar qq\rangle^2$ form. In our case
\ba
\alpha_s\langle (\bar u\gamma_{\alpha}\gamma_5 \tau^a u 
- \bar d\gamma_{\alpha}\gamma_5 \tau^a d)^2 \rangle |_{vacuum \, dominance} =
\nonumber \\ 
=\frac{32}{9}\alpha_s\langle \bar qq\rangle^2 \approx 6.5\cdot 10^{-4}\, {\rm GeV}^6  \\
\alpha_s\langle (\bar d\gamma_{\alpha} \tau^a d) \sum_{i=u,d,s} 
\langle  \bar q_i\gamma_{\alpha}\tau^a q_i  \rangle |_{vacuum \,dominance}=
\nonumber \\
= -\frac{32}{9}\alpha_s\langle \bar qq\rangle^2 \approx -6.5\cdot 10^{-4}\, {\rm GeV}^6,
\ea
with the numerical values following from the estimate
\be
\alpha_s\langle \bar qq\rangle^2 \approx -1.8\cdot 10^{-4} {\rm GeV}^6
\end{equation}
which can be justified using the above relation between the quark condensate, 
quark masses ($m_u+m_d=11$ MeV, experimental values of the  pion mass and
pion  decay constant $f_{\pi}$, under the assumption that  the relevant 
normalization point corresponds to $\alpha_s=0.7$. 
In fact, this estimate amounts to fixing one of the 
most important input parameters of the QCD sum rule
method and the quoted value is supported by successful predictions
for many different resonances.

\subsection{ Qualitative analysis of the QCD sum rule} 

Using the vacuum dominance
hypothesis, we can rewrite the sum rule in a more compact form:
\ba
\int_0^{\infty} e^{-s/M^2} R^{I=1}(s)ds = 
\frac3{2} M^2 \biggl [1+\frac{\alpha_s(M)}{\pi}-
\frac{2\pi^2f_{\pi}^2m_{\pi}^2}{M^4} +
\nonumber \\
+\frac{\pi^2}{3M^4}\langle \frac{\alpha_s}{\pi}GG \rangle -
\frac{448\pi^3}{M^6}\alpha_s \langle \bar qq\rangle^2 \biggr ].
\ea
Substituting the numerical values of the condensate parameters gives
\be
\int_0^{\infty}  e^{-s/M^2} R^{I=1}(s)ds = 
\frac3{2} M^2 \biggl [1+\frac{\alpha_s(M)}{\pi} 
\nonumber \\ 
+ 0.1 \left (\frac{0.6 {\rm GeV}^2}{M^2}\right )^2 - 
0.14 \left (\frac{0.6 {\rm GeV}^2}{M^2}\right )^3\biggr].
\end{equation}
Our experience with the  sum rules for the 
quantum-mechanical oscillator (see Section  2) 
suggests that to calculate the $\rho$-meson mass we should 
incorporate also the ``daughter'' sum rule resulting from the
original one after the differentiation with respect to $1/M^2$:
\be
\int_0^{\infty} e^{-s/M^2} sR^{I=1}(s)ds = 
\frac3{2} M^4 \biggl [1+\frac{\alpha_s(M)}{\pi} 
\nonumber \\ 
- 0.1(\frac{0.6 {\rm GeV}^2}{M^2})^2 +
 0.28 (\frac{0.6 {\rm GeV}^2}{M^2})^3\biggr].
\end{equation}
Note now that  power corrections in the sum rules above are small
for the values of the Borel parameter $M$ 
as low as $M^2=m_{\rho}^2=0.6 \, {\rm GeV}^2$:
\ba
\int_0^{\infty} e^{-s/m_{\rho}^2} R^{I=1}(s)ds = 
\frac3{2} m_{\rho}^2 [1+0.1+0.1-0.14], \\
\int_0^{\infty} e^{-s/m_{\rho}^2} sR^{I=1}(s)ds = \frac3{2} m_{\rho}^4 
[1+0.1-0.1+0.28].
\ea
At such a low value of $M^2$ one should expect that the integral
over the physical cross section is dominated by the lowest resonance,
{\em i.e.,} by the $\rho$-meson. Its contribution to $R^{I=1}(s)$,
in the limit of the vanishing width, can be written as
\be
R^{I=1}_{\rho}(s)=\frac{12\pi^2 m_{\rho}^2}{g_{\rho}^2}\delta(s-m_{\rho}^2).
\end{equation}
 
These observations allow one to make 
a rough estimate of the predictions following from the QCD 
sum rule. To this end we, following SVZ\cite{SVZ} 
 \\ $\bullet$  neglect, for $M^2=m_{\rho}^2$, the 
power corrections in the r.h.s. of the sum rules,
\\ $\bullet$  neglect, for  $M^2=m_{\rho}^2$, the higher state contributions in the
l.h.s.
This gives  the following relation  
\be
\frac{12\pi^2 m_{\rho}^2}{g_{\rho}^2}\, e^{-1} \approx \frac{3}{2} m_{\rho}^2.
\end{equation}
Simplifying it results in  a famous SVZ\cite{SVZ}  expression for the $g_{\rho}$
constant in terms of fundamental constants $\pi$ and $e$:
\be
\frac{g_{\rho}^2}{4\pi} \cong \frac{2\pi}{e} \approx 2.3,
\end{equation}
to be compared with the experimental value $2.36 \pm 0.18 $.

\subsection{Fitting the spectrum parameters}

The standard procedure is to extract the parameters of the 
hadronic spectrum requiring the best agreement between the
two sides of the sum rule. Of course, having only a few first terms
of the operator product expansion one should not hope to
reproduce all the details of the resonance structure in a
particular channel. Only the gross features of 
the spectrum can be extracted from the QCD  sum rules.
So, let us take the ansatz ``first narrow resonance plus continuum''
which has been  discussed in the oscillator case. In fact, one should expect that
in QCD this ansatz should work better, since the higher hadronic
resonances are very broad, and the total cross section of their production
is known to be well reproduced by the parton model ({\em i.e.}, perturbative)
calculation. Specifically, we take
\be
R^{I=1}_{\rho}(s)=\frac{12\pi^2 m_{\rho}^2}{g_{\rho}^2}\delta(s-m_{\rho}^2)
 +\frac3{2}\left(1+\frac{\alpha_s}{\pi}\right) \theta(s-s_0).
\end{equation}
The parameters to fit are the coupling constant $g_{\rho}^2$, the 
$\rho$-meson mass $m_{\rho}^2$ and the effective continuum threshold $s_0$.
 
The strategy will be just the same as  used for the oscillator states:
 \\ $\bullet$ {transfer the continuum contribution to the right hand side,}
\\ $\bullet$ {write the original sum rule in the form
\be
f_{\rho}^2e^{-m_{\rho}^2/M^2}=
\frac3{2}M^2\left(1-e^{-s_0/M^2}\right) + ``power\, corrections",
\end{equation}
where, for brevity, we introduced the notation 
$f_{\rho}^2=\frac{12\pi^2 m_{\rho}^2}{g_{\rho}^2}$,
\\ $\bullet$  write the daughter sum rule in the form
\be
f_{\rho}^2 m_{\rho}^2 e^{-m_{\rho}^2/M^2}=
\frac3{2}M^4\left(1-(1+\frac{s_0}{M^2})e^{-s_0/M^2}\right) 
+ ``power\, corrections",
\end{equation}
}
\\ $\bullet$ {divide the second sum rule by the first to get the expression 
for $m_{\rho}^2$ as a function of $s_0$ and $M^2$  (for fixed values
of  the condensates),}
\\ $\bullet$ {look for the value of the effective threshold $s_0$ that produces 
the most stable curve for $m_{\rho}^2$ (recall  that with all power 
corrections taken into account, and the exact ansatz for the higher states,
one should obtain an $M^2$-independent result for $m_{\rho}^2$).}

In this way one  would obtain\footnote{With all the numbers
explicitly given,  the readers are encouraged  
to perform  the fitting as an excercise.}
   $s_0 \approx 1.5 \, {\rm GeV}^2$ and 
$m_{\rho}^2 \approx 0.58 \, {\rm GeV}^2$.
Next step is to extract $g_{\rho}^2$ from the first sum rule, using these
values of $s_0$ and $m_{\rho}^2$.  The value 
$s_0 \approx 1.5 \, {\rm GeV}^2$  provides the most stable curve for 
$g_{\rho}^2$, corresponding to $g_{\rho}^2 \approx 2.17$. 
The result is 
rather sensitive to the $m_{\rho}^2$-value. If one takes the experimental
value $m_{\rho}^2 \approx 0.6 \, {\rm GeV}^2$, 
then the most stable curve 
still is that for $s_0 \approx 1.5 
{\rm GeV}^2$, but the $g_{\rho}^2$-value is
a little higher  $g_{\rho}^2 \approx 2.40$ and closer to the experimental
one $g_{\rho}^2 \approx 2.36 \pm 0.18$. Anyway, it should be emphasized
that all the $g_{\rho}^2$-values extracted from the QCD sum rule differ
from the experimental value by less than $10\%$, the claimed 
 precision of the method.

\section{ QCD Sum Rules for  Pion}

\subsection{ Axial vector current}

 Vector currents are a special
case, because the relevant correlators can be related to directly
measurable cross sections. From purely theoretical point of view,
other currents such as scalar, pseudoscalar, axial vector {etc.,} 
are not worse, but experimental information in these cases is 
normally limited to the masses of the lowest states. Still, the 
axial vector current is in a better situation. First, its projection
onto the pion state is just the pion decay constant $f_{\pi}$:
\be
\langle 0 | \bar d \gamma_5 \gamma_{\mu} u |P \rangle = i f_{\pi} P_{\mu}.
\end{equation}
Second, measuring the decays of the heavy lepton $\tau \to \nu_{\tau} + X$,
one can measure the coupling of a hadronic state $X$ to the axial 
current. Our goal now is to show how one can use the QCD sum rules
to study the hadronic spectrum in the axial vector channel.
Again, our presentation here closely follows the classic 
SVZ analysis\cite{SVZ}.

Since the axial vector current $a_{\mu}(x)=\bar d \gamma_5 \gamma_{\mu} u$
 is not conserved, the relevant 
correlator $\Pi^{(ax)}_{\mu\nu}(q)$ is a sum of two independent functions:
\be
\Pi^{(ax)}_{\mu\nu}(q) =
i \int e^{iqx}  \langle 0 | T \,(a_{\mu}^+(x)\,a_{\nu}^-(0)
\,)|\,0\rangle\, d^4 x = 
-g_{\mu\nu}\Pi_1(Q^2)+q_{\mu}q_{\nu}\Pi_2(Q^2),
\end{equation}
or, in terms of the transverse and longitudinal components
\be
\Pi^{(ax)}_{\mu\nu}(q) = -\left 
\{g_{\mu\nu}-\frac{q_{\mu}q_{\nu}}{q^2}\right \}\Pi_1(Q^2)-
\frac{q_{\mu}q_{\nu}}{q^2}\{\Pi_1(Q^2)+Q^2\Pi_2(Q^2) \}.
\end{equation}

\subsection{ Massless pion}

The common belief is that the pion mass vanishes in the limit
of exact chiral symmetry so that the pion is a Goldstone particle.
It can be shown, that the operator product expansion in this case
really requires the presence of a massless particle in the limit $m_q \to 0$.
In this limit the imaginary part of the longitudinal function 
$\Pi_1(Q^2)+Q^2\Pi_2(Q^2)\equiv \Pi_{\mid\mid}$ vanishes, and 
$\Pi_{\mid\mid}$ is a polynomial in $Q^2$. Now, let us switch on
a small quark mass and keep terms linear in this mass. 
The statement is that the function $\Pi_1(Q^2)+Q^2\Pi_2(Q^2)$ is 
exactly calculable in this approximation:
\be
\Pi_1(Q^2)+Q^2\Pi_2(Q^2)=
\frac{(m_u+m_d)\langle \bar uu + \bar dd \rangle}{Q^2} + O(m_q^2).
\end{equation}
There are no terms linear in $m_q$ 
of higher orders in $1/Q^2$. Let us show that
this equation really implies the existence of a nearly massless pion.
First we should express the $\Pi$-functions in terms of 
the hadronic contributions, using the dispersion relations:
\ba
\Pi_1(Q^2)= \frac1{\pi}\int_0^{\infty}\frac{{\rm Im}\, \Pi_1}{s+Q^2}ds + ``subtr."\\
\Pi_2(Q^2)= \frac1{\pi}\int_0^{\infty}\frac{{\rm Im}\, \Pi_2}{s+Q^2}ds + ``subtr."\\
\Pi_1(Q^2)+Q^2\Pi_2(Q^2)= 
 \frac1{\pi}\int_0^{\infty}\frac{{\rm Im}\, 
 \Pi_1-s{\rm Im}\,\Pi_2}{s+Q^2}ds + ``subtr."
\ea
where ${\rm Im}\, \Pi_1$ contains contributions only from pseudovector 
(spin-1) states and 
their contribution should be transverse,
{\rm i.e.,} proportional to $(g_{\mu\nu}-q_{\mu}q_{\nu}/q^2)$:
\be
{\rm Im}\, \Pi_1 = \sum_{A} \pi f_A^2\delta(s-M^2),
\end{equation}
while  ${\rm Im} \, \Pi_2$ contains contributions both from 
pseudovector (spin-1) states and from the pseudoscalar (spin-0) ones
and their contribution into $\Pi^{(ax)}_{\mu\nu}(q)$ should be longitudinal
{\em i.e.,} proportional to $q_{\mu}q_{\nu}$:
\be
s{\rm Im} \Pi_2 = \sum_{A} \pi f_A^2\delta(s-M_A^2) +
\sum_{P} \pi f_P^2m_P^2\delta(s-M_P^2).
\end{equation}
Hence, only pseudoscalar states contribute to the longitudinal
component of $\Pi^{(ax)}_{\mu\nu}$. For small quark masses we have
\ba
\frac{(m_u+m_d)\langle \bar uu + \bar dd \rangle}{Q^2} + O(m_q^2)= 
-\sum_P \frac{f_P^2m_P^2}{Q^2+m_P^2}= \nonumber\\
=-\sum_P \left (\frac{f_P^2m_P^2}{Q^2}-\frac{f_P^2m_P^4}{Q^4}+\ldots \right )
\  . 
\ea
All $O(Q^{-4})$ terms should be $O(m_q^2)$. This means that
\begin{equation}
f_P^2m_P^4=O(m_q^2)
\end{equation} 
for all states. Hence, the particles with masses 
that remain constant as $m_q \to 0$, 
decouple in the chiral limit:
\ba
m_P^2=O(m_q^0),\\
f_P^2=O(m_q^2).
\ea
The last relation also means that they do not contribute 
to the term linear in $m_q$. The only way to get this term 
is to assume that there exists a state for which the $O(m_q)$-term
is brought by the $m_P^2$-factor:
\ba
m_P^2=O(m_q^1),\\
f_P=O(m_q^0).
\ea
Such a state is naturally identified with the pion. In addition, we get
the well-known relation
\be
f_{\pi}^2m_{\pi}^2 = - (m_u+m_d)\langle \bar uu + \bar dd \rangle
\end{equation}
demonstrating explicitly the vanishing of the pion mass in the chiral
limit.

To derive the starting statement, one should consider 
$q_{\mu}q_{\nu}\Pi^{(ax)}_{\mu\nu}(q)$. By equations of motion
it is related to the correlator of the pseudoscalar densities:
\ba
q_{\mu}q_{\nu}\Pi^{(ax)}_{\mu\nu}(q)=Q^2(\Pi_1(Q^2)+Q^2\Pi_2(Q^2))= \nonumber \\
= -i \int e^{iqx}  \langle 0 | T \,
\left((\bar u(x) \gamma_5 d(x))\ (\bar d(0) \gamma_5  u(0))
\,\right)|\,0\rangle\, d^4x +{\rm const.}
\ea
The constant on the right-hand side is due to contact terms which normally 
arise if one differentiates a $T$-product. This constant corresponds to the
$1/Q^2$-term in the combination $\Pi_1(Q^2)+Q^2\Pi_2(Q^2)$. Direct calculation
of this term gives the result stated in the beginning of this subsection.

Thus, the operator product expansion implies that $m_{\pi}^2 \to 0$ 
in the $m_q \to 0$ limit, and , hence, the pion mass should not
be fitted from the sum rule. Within our accuracy one can set 
$m_{\pi}^2 = 0$ from the start.

\subsection{ QCD sum rule for the axial channel} 

The operator product 
expansion for the correlator of two axial currents is completely
analogous to that of two vector currents, the  only difference being
the presence of the extra $\gamma_5$-matrices in the vertices
corresponding to the currents.  For massless quarks, however, this
produces no changes both in perturbative and $O(\langle GG \rangle)$
contributions. However, the four-quark terms 
shown in  Fig.\ref{fig:rocond}e,f 
change sign, because one
of the $\gamma$-matrix type terms is substituted by the condensate
factor $\langle \bar qq \rangle$ which has the structure of {\bf 1}
with respect to the Dirac indices.

After the Borel transformation the sum rule looks as follows:
\ba
\int_0^{\infty} e^{-s/M^2} {\rm Im}\,\Pi_2(s)ds = 
\frac{M^2}{4\pi} \biggl [1+\frac{\alpha_s(M)}{\pi}+
\frac{\pi^2}{3M^4}\langle \frac{\alpha_s}{\pi}GG \rangle +
 \nonumber \\
+\frac{4\pi^3}{M^6}\langle \alpha_s(\bar u\gamma_{\alpha}\gamma_5 \tau^a d 
\, \bar d\gamma_{\alpha}\gamma_5 \tau^a u) \rangle +
\nonumber \\
+ \frac{4\pi^3}{9 M^6} \langle \alpha_s (\bar u\gamma_{\alpha}\tau^a u 
+ \bar d\gamma_{\alpha} \tau^a d) \sum_{q=u,d,s} 
\langle 0 | \bar q\gamma_{\mu}\tau^a q |0 \rangle \biggr] \, .
\ea

In the axial channel we have two states that are the lowest
among other states with the same quantum numbers: a pseudoscalar (the pion)
and an axial vector ($A_1$-meson) ones. Thus, one can try the ansatz with
two low-lying states (``pion+$A_1$+continuum''):
\be
{\rm Im}\,\Pi_2(s) = \pi f_{\pi}^2 \delta(s) + 
\pi f_{A_1}^2 \delta(s-m_{A_1}^2) +
\frac1{4\pi}\left(1+\frac{\alpha_s}{\pi}\right) \theta(s>s_0^{(A_1)}).
\end{equation}
One can also treat the $A_1$ as a part of the continuum. This, 
more crude approximation, corresponds to the simplified ansatz
\be
{\rm Im}\,\Pi_2(s) = \pi f_{\pi}^2 \delta(s) + 
\frac1{4\pi}\left(1+\frac{\alpha_s}{\pi}\right) \theta(s>s_0^{(\pi)}).
\end{equation}
The constants $f_{\pi}, f_{A_1}$ are defined as follows:
\be
\langle 0 | \bar d \gamma_5 \gamma_{\mu} u |\pi \rangle = i f_{\pi} P_{\mu},
\hspace{1cm}
\langle 0 | \bar d \gamma_5 \gamma_{\mu} u |A_1 \rangle = 
i f_{A_1} \epsilon_{\mu},
\end{equation}
where $\epsilon_{\mu}$ is the $A_1$ polarization vector.

\subsection{ Fitting the sum rule} 

Substituting the numerical values
of the condensates into the basic sum rule gives
\ba
\int_0^{\infty}  e^{-s/M^2} {\rm Im}\,\Pi_2(s)ds = 
\frac{M^2}{4\pi} \biggl [1+\frac{\alpha_s(M)}{\pi} 
\nonumber \\ 
+ 0.1 \left (\frac{0.6 {\rm GeV}^2}{M^2} \right)^2 + 
0.22 \left (\frac{0.6 {\rm GeV}^2}{M^2}\right )^3 \biggr ].
\ea
The ``daughter'' sum rule is 
\ba
\int_0^{\infty}  e^{-s/M^2} {\rm Im}\,\Pi_2(s)sds = 
\frac{M^2}{4\pi}\biggl [1+\frac{\alpha_s(M)}{\pi} 
\nonumber \\ 
- 0.1\left (\frac{0.6 {\rm GeV}^2}{M^2}\right )^2 - 
0.44 \left (\frac{0.6 {\rm GeV}^2}{M^2}\right )^3
\biggr ].
\ea
Note, that in the $m_{\pi}^2=0$ approximation
 the pion does not contribute to the daughter sum rule, and the
spectrum structure for $s{\rm Im}\,\Pi_2(s)={\rm Im}\,\Pi_1(s)$
is more like in the vector channel: first non-zero mass resonance $+$
continuum. This is reflected in the structure of the relevant sum rule:
power corrections are negative. They are considerably larger than those 
in the $\rho$-channel, and one should expect that the $A_1$-mass (squared) 
is much larger than $m_{\rho}^2$.

The fitting procedure \footnote{Again, we encourage
the readers to obtain the curves themselves.} 
 is most straightforward 
in the case of the simplified ansatz. One should transfer the continuum 
contribution to the right hand side to get $f_{\pi}^2$ as a
function of $s_0^{(\pi)}$ and the Borel parameter $M^2$:
\be
4\pi^2 f_{\pi}^2 = 
 M^2 \left(1-e^{-s_0^{(\pi)}/M^2}\right) \left [ 1+\frac{\alpha_s(M)}{\pi}\right ]
+\frac{\pi^2}{3M^2}\langle \frac{\alpha_s}{\pi}GG \rangle +
\frac{704\pi^3}{M^4}\alpha_s \langle \bar qq\rangle^2 \ .
\end{equation}
The most stable curve corresponds 
to $s_0 \approx 0.7 \, {\rm GeV}^2$
and  $4\pi^2 f_{\pi}^2 \approx 0.65 \, {\rm GeV}^2$, the experimental \
value being  $4\pi^2 f_{\pi}^2 = 0.67 \, {\rm GeV}^2$.

One can also try  the  ansatz in which $A_1$ is 
treated  as a (narrow) resonance. The strategy is just the same
as in the $\rho$-case. To this end, one should 
  \\ $\bullet$  differentiate the 
daughter (second) sum rule with respect to $1/M^2$ to get the third
sum rule,
\\ $\bullet$  transfer the continuum contribution to the right hand side
in all sum rules,
\\ $\bullet$  extract $m_{A_1}^2$ and $s_0^{(A_1)}$ from the ratio of the third 
sum rule to the second, 
\\ $\bullet$   using the values $m_{A_1}^2 \approx 1.6 \, {\rm GeV}^2$ 
and $s_0^{(A_1)} \approx 2.4\, {\rm GeV}^2$
obtained from the previous fitting,
find $f_{A_1}^2$ from the second sum rule,
\\ $\bullet$ 
using the value $4\pi^2f_{A_1}^2 \approx 1.95 \, {\rm GeV}^2$ 
obtained from the previous fitting,
find $f_{\pi}^2$ from the first sum rule.
The value $f_{\pi}^2 \approx 0.69 \, {\rm GeV}^2$ obtained in this way
is very close to the experimental one $f_{\pi}^2|^{exp}=0.67 \, {\rm GeV}^2$.
The $A_1$ mass value $m_{A_1}^2 \approx 1.6 \, {\rm GeV}^2$ 
extracted from the QCD sum rule also is in 
a very good agreement with experiment.

\section{ QCD Sum Rules for   Pion Form Factor}

\subsection{ Three-point function} 

To analyze the pion form factor,
we  consider the three-point function, {\em i.e.,} the 
correlator of the three currents\cite{ios,ffsr}:
\be
T^{\mu}_{\alpha\beta}(p_1,p_2) =
i \int e^{-ip_1x+ip_2y} 
\langle \{j_{\beta}(y) J^{\mu}(0) j^+_{\alpha}(x)\}\rangle d^4x d^4y ,
\end{equation}
where 
\be
J^{\mu} = \frac2{3}\bar u \gamma^{\mu} u - \frac1{3} \bar d \gamma^{\mu} d
\end{equation}
is the electromagnetic current and 
\be
j_{\alpha} = \bar d \gamma_5 \gamma_{\alpha} u
\end{equation}
is the axial current  used in the preceding section to study
the static properties of the pion. The pion contribution 
\be
\langle 0|j_{\beta}(y)|p_2\rangle \langle p_2 |J^{\mu}(0)|p_1 \rangle 
\langle p_1| j^+_{\alpha}(x)| 0 \rangle 
\end{equation}
into this three-point correlator can be obtained by inserting the 
pion state(s) between the currents. The form factor $F_{\pi}(q^2)$ 
appears in the middle matrix element:
\be
\langle p_2 |J^{\mu}(0)|p_1 \rangle = (p_1^{\mu}+p_2^{\mu}) F_{\pi}(q^2),
\end{equation}
while the other two are projections of the axial current onto the 
initial and final pion states
\ba
\langle 0|j_{\beta}(0)|p_2\rangle = i f_{\pi} (p_2)_{\beta}\\
\langle p_1| j^+_{\alpha}(0)| 0 \rangle = i f_{\pi} (p_1)_{\alpha}.
\ea

The three-point function $T^{\mu}_{\alpha\beta}(p_1,p_2)$ 
is the sum of different structures each characterized by the
relevant invariant amplitude $t_i(p_1^2,p_2^2,q^2)$. The first
idea is to study the pion form factor analyzing the invariant
amplitude corresponding to the structure 
$(p_2)_{\beta}(p_1)_{\alpha}(p_1^{\mu}+p_2^{\mu})$. However, 
there are other structures
($(p_1)_{\beta}(p_2)_{\alpha}, (p_1)_{\beta}(p_1)_{\alpha} ,
(p_2)_{\beta}(p_2)_{\alpha}$ that  coincide with 
$(p_2)_{\beta}(p_1)_{\alpha}$ in the 
limit $q \to 0$. This complication disappears if all the basic structures 
are expanded in $P=(p_1+p_2)/2$ and $q=p_2-p_1$. Then the 
pion contribution is
$$
(p_2)_{\beta}(p_1)_{\alpha}(p_1^{\mu}+p_2^{\mu})= 
2\left (P_{\beta}+\frac{q_{\beta}}{2}\right )
\left (P_{\alpha}-\frac{q_{\alpha}}{2}\right )P^{\mu}=
2P_{\beta}P_{\alpha}P^{\mu}+ O(q),
$$
and the ``best'' structure (not changing when $q$ varies) is 
$P_{\alpha}P_{\beta}P^{\mu}$. The simplest way to extract 
the relevant amplitude (hereafter referred to as $T$) is to multiply
$T^{\mu}_{\alpha\beta}(p_1,p_2)$  by $n_{\mu}n^{\alpha}n^{\beta}$,
where $n$ is a lightlike vector 
orthogonal to $q$.
The property $n^2=0$ kills all
the structures containing $g_{\alpha\beta}$ and other $g$-terms, while
 the property $(nq)=0$ kills all the structures
containing any $q$-factor.  Thus, the amplitude we are going to 
analyze is
\be
T(p_1^2,p_2^2,q^2)=\frac{n_{\mu}n^{\alpha}n^\beta}{2(nP)^3}
 T^{\mu}_{\alpha\beta}(p_1,p_2).
\end{equation}

\subsection{ Dispersion representation}

To extract information about the form factors of physical states,
one should incorporate the double dispersion relation
\be
T(p_1^2,p_2^2,q^2)=\frac1{\pi^2}\int_0^\infty ds_1\int_0^\infty ds_2
 \ \frac{\rho(s_1,s_2,q^2)}{(s_1-p_1^2)(s_2-p_2^2)}
+ ``subtractions". \label{eq:drn}
\end{equation}
The subtraction terms, in general, are polynomial in $p_1^2$ and/or $p_2^2$.
They disappear after one applies 
$B(p_1^2\to M_1^2)B(p_2^2\to M_2^2)$, the Borel transformation  
in both variables \footnote{ In our specific case, the subtractions 
are exhausted by a constant, and a single 
differentiation with respect to any variable is sufficient to remove it.}.
The double Borel transform $\Phi=B_1B_2T$ is related to the
double spectral density $\rho(s_1,s_2,q^2=-Q^2)$ by
\be
\Phi(M_1^2,M_2^2,q^2=-Q^2)=\frac1{\pi^2}\int_0^\infty \frac{ds_1}{M_1^2}
\int_0^\infty \frac{ds_2}{M_2^2} 
\exp \left (-\frac{s_1}{M_1^2}-\frac{s_2}{M_2^2}\right )
\rho(s_1,s_2,Q^2).
\end{equation}

\subsection{Operator product expansion}

  \begin{figure}[t]
\mbox{
   \epsfxsize=12cm
 \epsfysize=3cm
  \epsffile{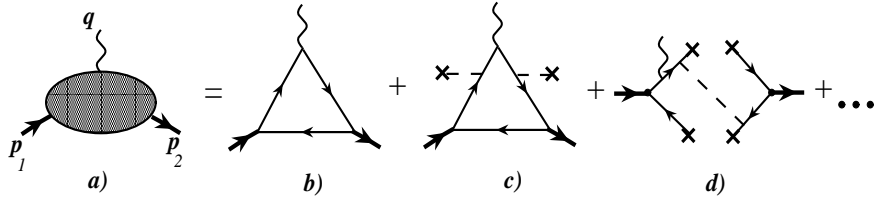}  }
{\caption{\label{fig:3point} 
Structure of OPE in case of symmetric kinematics.  
}}
\end{figure}

In contrast to the two-point
correlators $\Pi(Q^2)$ analyzed in the previous Section, the three-point
amplitudes $T(p_1^2,p_2^2,q^2)$ have three arguments. To incorporate
asymptotic freedom, one should take the invariants $p_1^2,p_2^2$
corresponding to the ``hadronized'' channels to be negative and
sufficiently large, say, $|p_1^2|,|p_2^2|>1 \, {\rm GeV}^2$. Depending on the
value of the third invariant $q^2$  
that works as an external parameter (it also should be negative,
of course), there are three 
essentially different situations.
 \\ $\bullet$  The simplest is the symmetric case when $Q^2$ is of the
same order of magnitude as $|p_1^2|,|p_2^2|$:
$$|p_1^2|\sim|p_2^2|\sim Q^2\sim \mu^2.$$
 The operator product expansion for $T$ in this case has the 
following structure:
\be
T(p_1^2,p_2^2,q^2)= \sum_{i}\frac1{(\mu^2)^{d_i}} 
C_i \left (\frac{\mu^2}{p_1^2},\frac{\mu^2}{p_2^2},
\frac{\mu^2}{q^2} \right )
\langle {\cal O}_i \rangle
\end{equation}
where $C_i/(\mu^2)^{d_i}$ is a perturbatively calculable short-distance
coefficient function, $\langle {\cal O}_i \rangle$ is the vacuum
matrix element of a local operator , {\em i.e.,} some condensate
term  and  $d_i$ is the dimension of ${\cal O}_i$ in mass units
(see Fig.\ref{fig:3point}).
\\ $\bullet$    The case when $Q^2$ is small is more complicated: 
$$|p_1^2|\sim|p_2^2|\sim \mu^2,\hspace{1cm}Q^2 \ll \mu^2.$$
The coefficient functions $C_i$ calculated in  the symmetric
kinematics, might become singular in this limit because of $(\mu^2/Q^2)^n$-
 or $\log(\mu^2/Q^2)$-terms. The operator product expansion 
in such a situation has a modified form:
\ba
T(p_1^2,p_2^2,q^2)= \sum_{i}\frac1{(\mu^2)^{d_i}} 
\tilde C_i \left (\frac{\mu^2}{p_1^2},\frac{\mu^2}
{p_2^2},\frac{\mu^2}{q^2} \right
)
\langle {\cal O}_i \rangle \nonumber \\ + \sum_{j} \frac1{(\mu^2)^{d_j}} 
 E_k(\frac{\mu^2}{p_1^2},\frac{\mu^2}{p_2^2})
\Pi_k(Q^2)
\ea
where the coefficient functions $\tilde C_i$ are regular in the $Q^2 \to 0$
limit, $\Pi_k(Q^2)$ is the correlator of the
electromagnetic current $J$ and a local operator ${\cal O}_k$ 
$$
\Pi_k(Q^2) = i \int e^{-iqx} \langle {\cal O}_k(x)\, J(0) \rangle\,d^4x,
$$
$d_k$ is the dimension of ${\cal O}_k$ in mass units. The coefficient
functions $E_k$ are (the Fourier transforms of) 
those for the operator product expansion of two
axial currents 
$$
T(j\,j^+)=\sum_{k} E_k \, {\cal O}_k.
$$

In the modified OPE for the three-point function,
the contributions of the first type correspond to a situation
when large momentum flow connects all three external vertices, 
while those of the second type describe the situation when large momentum
flows only between the vertices $x$ and $y$ related to the axial currents.
The momenta in the $J$-channel are of the order of $Q$, {\em i.e.,} small,
and the correlator $\Pi_j(Q^2)$ is a nonperturbative object accumulating
long-distance information. 
\\ $\bullet$  In the opposite limit of large $Q^2$
$$|p_1^2|\sim|p_2^2|\sim \mu^2,\hspace{1cm}Q^2\gg \mu^2,$$
a simple expansion in powers of 
$1/|p_1^2|, 1/|p_2^2$ ({\em i.e.,}in $1/\mu^2$)
might be destabilized by large $(Q^2/\mu^2)^n$-terms. Appearance of
terms growing with $Q^2$ is a bit surprizing, since form factors
should decrease as $Q^2 \to \infty$. But the appearance of such terms
is just the artifact of the $1/\mu^2$ expansion: even a very reasonable
function like $1/(Q^2+\mu^2)$, vanishing as $Q^2 \to \infty$, produces
a divergent series when expanded in $1/\mu^2$. Thus, a possible way
out in the large-$Q^2$ limit is a resummation of the parametrically
enhanced contributions of $(Q^2/\mu^2)^n$-type. In fact, it can be established,
that such contributions result from the Taylor expansion of the 
original nonlocal combinations 
$\langle \varphi (0) \varphi (x) \rangle$ (nonlocal condensates), 
and the idea is to construct sum rules without expanding the nonlocal
condensates, treating them as some phenomenological functions
describing the propagation of quark and gluons in the
physical vacuum.

\subsection{Perturbative contribution} 

The structure of the 
power corrections in the operator product expansion, as we discussed
above, depends on the particular $Q^2$-region under study. However,
the starting point of any expansion is the perturbative contribution, 
in our case corresponding to the triangle diagram plus radiative 
corrections to it.

Using Feynman parametrization, one can rather easily obtain
the following representation\cite{ffsr}  for the double Borel transform
of the lowest-order perturbative triangle diagram 
(see Fig.\ref{fig:3point}a) 
\ba
\Phi^{pert}(M_1^2,M_2^2,Q^2) = \frac3{2\pi^2(M_1^2+M_2^2)}\,
\int_0^1 \, x(1-x) \nonumber \\
\exp\left\{-\frac{xQ^2}{(1-x)(M_1^2+M_2^2)}\right\}\, 
I(x,M_1^2,M_2^2,m_q^2,e_q) \, dx
\ea
where $I(x,M_1^2,M_2^2,m_q^2,e_q)$ is the factor containing the dependence
on the quark masses
\be
I(x,M_1^2,M_2^2,m_q^2,e_q)= e_u \exp\left\{-\left(\frac{m_u^2}{1-x}
+\frac{m_d^2}{x}\right)
\left(\frac1{M_1^2}+\frac1{M_2^2}\right)\right\} + \{u \leftrightarrow d \}.
\end{equation}
In view of extreme smallness of the $u$- and $d$-quark masses, this
factor can be safely set to unity. The $x$-variable may be interpreted
as the fraction of the total momentum $P$ carried by the passive quark 
(active is the quark interacting with the $EM$ current).
This integral representation is very convenient to study 
the behavior of $\Phi^{pert}(M_1^2,M_2^2,Q^2)$
for small and large $Q^2$.
%
In the low-$Q^2$ limit we get
\ba
&&\lefteqn{ \Phi^{pert}(M_1^2,M_2^2,Q^2)  =  \frac1{4\pi^2(M_1^2+M_2^2)}\biggl\{1 - 
2\frac{Q^2}{M_1^2+M_2^2} } \\
&& -\sum_{N=0}^{\infty} 
\left(\frac{-Q^2}{M_1^2+M_2^2}\right)^{N+2}\frac{N+3}{N!}
\left(\log\left[\frac{Q^2}{M_1^2+M_2^2}\right] + \frac1{N+3} - 
\psi(N+1)\right)\biggr\} \ . \nonumber 
\ea
The first two terms of the expansion can be obtained by simply 
expanding the exponential factor in $Q^2$ under the integral sign.
However, the third term of such an expansion produces a logarithmically
divergent integral
$$
\ldots \int_0^1 \frac1{2!}\left(\frac{Q^2}{M_1^2+M_2^2}\right)^2
\frac{x^3}{1-x} \,dx ,
$$
and higher terms have power divergences when integrated in the region 
$x \to 1$. This region corresponds to a  situation when the whole
large momentum $P$ is carried by the passive quark, and the active
quark carries small momentum proportional to $q$. 
This is precisely the situation producing additional terms
in the operator product expansion at small $Q^2$. 
The logarithmic  factor $\log(Q^2)$,
singular when $Q^2 \to 0$,  just signalizes that there are
long-distance effects (quarks propagate over 
distances $\sim 1/Q$) limiting the applicability of perturbation theory.

In the large-$Q^2$ limit one should expand 
$\Phi^{pert}(M_1^2,M_2^2,Q^2)$ in $1/Q^2$ rather
than in $Q^2$:
\ba
\Phi^{pert}(M_1^2,M_2^2,Q^2) = \frac3{2\pi^2}\frac{M_1^2+M_2^2}{(Q^2)^2}
\biggl\{1 - 8\frac{M_1^2+M_2^2}{Q^2}+
\nonumber \\ +
\frac1{6}\sum_{N=2}^{\infty} \,
(-)^N (N+1)(N+3)!\left(\frac{M_1^2+M_2^2}{Q^2}\right)^N \biggr\} .
\ea
Thus, the perturbative contribution vanishes like $1/Q^4$ as $Q^2 \to \infty$. 
Combining the spectral representation for  
$\Phi(M_1^2,M_2^2,Q^2)$ and the explicit
expression for it, we get the equation for  the perturbative
spectral density $\rho^{pert}(s_1,s_2,Q^2)$ 
\ba
\frac1{\pi^2}\int_0^\infty \frac{ds_1}{M_1^2}
\int_0^\infty \frac{ds_2}{M_2^2} \exp(-\frac{s_1}{M_1^2}-\frac{s_2}{M_2^2})
\rho^{pert}(s_1,s_2,Q^2) = \nonumber \\
= \frac3{2\pi^2(M_1^2+M_2^2)}
\int_0^1 \, x(1-x) \exp\left(-\frac{xQ^2}{(1-x)(M_1^2+M_2^2)}\right)  \, dx .
\ea
Solving it (this amounts to a double inverse Laplace transformation) gives
the perturbative spectral density:
\be
\rho^{pert}(s_1,s_2,Q^2)=\frac{3}{2} \left[ \frac{(Q^2)^2}{2!} +
\frac{(Q^2)^3}{3!} \right]\frac1{\sqrt{(s_1+s_2+Q^2)^2-4s_1s_2}}.
\end{equation}
In the low-$Q^2$  limit the perturbative spectral density 
can be expanded in $Q^2$:
\ba
\rho^{pert}(s_1,s_2,Q^2)|_{Q^2\to 0} = \frac1{4} \delta(s_1-s_2)\theta(s_1)
\theta(s_2)+\nonumber \\
+\frac{Q^2}{4}(s_1+s_2) \delta^{\prime\prime}(s_1-s_2)\theta(s_1)
\theta(s_2)+\ldots .
\ea
For large $Q^2$ one can expand it in $1/Q^2$:
\be
\rho^{pert}(s_1,s_2,Q^2)|_{Q^2\to \infty} = 
\frac3{2} \, \theta(s_1) \, 
\theta(s_2)\left\{\frac{s_1+s_2}{Q^4}- 4\frac{s_1^2+s_2^2+4s_1s_2}{Q^6}+
\ldots\right\} .
\end{equation}
Thus, for small $Q^2$, the  perturbative spectral density 
is concentrated on the line $s_1=s_2$ : there are no transitions between
states with different masses. Along this line, the spectral density
does not vary: it has no dependence on $s_1+s_2$. 
In the opposite limit, when $Q^2 \to \infty$, the 
situation is, in a sense, reversed: the density is nonzero
in the whole region $s_1>0, s_2>0$, there is no dependence on 
$(s_1-s_2)$, and the density grows like $s_1+s_2$. The last observation 
 means that higher states are relatively 
more important for large $Q^2$ than
for the small ones.

\subsection{Sum rule for the pion form factor at moderate $Q^2$}

The physical spectral density $\rho(s_1,s_2,Q^2)$, of course, differs
from its perturbative analog. It  has a rather
complicated structure on the $s_1,s_2$-plane. In particular,
$\rho(s_1,s_2,Q^2)$ contains the term corresponding to the pion 
form factor
\be
\rho_{\pi\pi}(s_1,s_2,Q^2)= \pi^2f_{\pi}^2F_{\pi}(Q^2)
\delta(s_1-m_{\pi}^2)\delta(s_2-m_{\pi}^2).
\end{equation}
In addition, it contains the contributions corresponding to transitions 
between the pion and higher resonances, and also the terms related to
elastic and transition form factors of the higher resonances. 
Constructing the two-dimensional ($s_1,s_2$) analog of the 
simplest ansatz ``lowest state plus continuum''  we can treat
all the contributions, except for the $\rho_{\pi\pi}$, as ``continuum'',
{\em i.e.}, assume that $\rho(s_1,s_2,Q^2) = \rho^{pert}(s_1,s_2,Q^2)$ 
outside the square $(0,s_0)\times (0,s_0)$:
\be
\rho(s_1,s_2,Q^2) = \rho_{\pi\pi}(s_1,s_2,Q^2) + 
\left(1- \theta(s_1<s_0)\theta(s_2<s_0)\right)\rho^{pert}(s_1,s_2,Q^2).
\end{equation}
Now, transferring the continuum contribution to the right-hand side
of the sum rule and using the operator product expansion in the symmetric 
kinematics ($Q^2 \sim M_1^2, M_2^2$), we obtain the QCD sum rule for the pion
form factor:
\ba 
f_{\pi}^2F_{\pi}(Q^2) = \frac1{\pi^2}\int_0^{s_0}ds_1
\int_0^{s_0} ds_2 \exp\left(-\frac{s_1+s_2}{M^2}\right)\rho^{pert}
(s_1,s_2,Q^2)+
\nonumber \\ 
+\frac{\alpha_s\langle GG \rangle}{12\pi M^2} +
\frac{16}{81}\frac{\pi\alpha_s \langle \bar qq\rangle^2}{M^4}
\left(13+2\frac{Q^2}{M^2}\right).
\ea
To treat initial and final states on equal footing and to simplify
the analysis, we took $M_1^2=M_2^2$. Unfortunately,
the double integral 
in this sum rule cannot be calculated explicitly for arbitrary $Q^2$. 
However, it is 
instructive to consider the formal limit $Q^2=0$:
\be
f_{\pi}^2F_{\pi}(0) = \frac1{4\pi^2} \frac {M^2}{2}
\left(1-e^{-2s_0/M^2}\right) 
+\frac{\alpha_s\langle GG \rangle}{12\pi M^2} +
\frac{208}{81}\frac{\pi\alpha_s \langle \bar qq\rangle^2}{M^4}.
\end{equation}
This should be compared to the $f_{\pi}^2$ sum rule:
\be
f_{\pi}^2 = \frac{m^2}{4\pi^2} 
\left(1-e^{-s_0/m^2}\right) 
+\frac{\alpha_s\langle GG \rangle}{12\pi m^2} +
\frac{176}{81}\frac{\pi\alpha_s \langle \bar qq\rangle^2}{m^4}.
\end{equation}
The condensate terms look almost identical, but comparing the 
perturbative terms we observe that $M^2$, the Borel parameter
of the form factor sum rule, should be larger by factor 2 than 
$m^2$, the Borel parameter of the $f_{\pi}^2$ sum rule: $M^2=2m^2$. After this
change, the $\langle GG \rangle$-term in the form factor sum rule is 
multiplied by $\frac1{2}$ and  the $\langle \bar qq\rangle^2$-term 
is multiplied by $\frac1{4}$
\begin{equation}
f_{\pi}^2F_{\pi}(0) = \frac{m^2}{4\pi^2} 
\left(1-e^{-s_0/m^2}\right) 
+\frac{\alpha_s\langle GG \rangle}{24\pi m^2} +
\frac{52}{81}\frac{\pi\alpha_s \langle \bar qq\rangle^2}{m^4}.
\end{equation}
Thus, we obtained a sum rule, which looks like that for $f_{\pi}^2$, 
but with the gluonic condensate term smaller by factor $2\approx (1.4)^2$, and
the quark condensate term smaller by factor $\sim 3.5\approx (1.5)^3$. 
Since it is the values of the condensates that (after the fitting 
procedure) determine all  the 
hadronic scales, we conclude that all the hadronic paramerters
having the dimension of $(mass)^2$, {\em i.e.,} $s_0$ and the
combination 
$f_{\pi}^2F_{\pi}(0)$, extracted from this sum rule are by a factor
$1.4 \,- 1.5 $ smaller than $s_0^{\pi}$ and $f_{\pi}^2$, 
respectively\footnote{One should not worry about the fact that 
a straightforward extrapolation of the form factor sum rule  
to the point $Q^2=0$, 
gives $F_{\pi}(0)\approx 0.7$. This sum rule is valid only
for sufficiently high $Q^2$, and, as we emphasized, 
there appear additional terms in the OPE
which will increase the condensate contributions to make them 
equal to those in the $f_{\pi}^2$ sum rule and recover the $F_{\pi}(0)=1$
normalization condition.}.
To get $s_0$ closer to $s_0^{\pi}$, one should restore the 
matching between the perturbative and 
the condensate terms, {\em i.e.}, decrease the perturbative
term by a factor $2$ to $4$. This allows us to make
an educated guess that  QCD sum rule for the pion 
form factor will most effectively work in the intermediate region
where the form factor varies between 
$0.5$ and $0.25$,
{\em i.e.,} for $Q^2$ between $0.5 \, {\rm GeV}^2$ and $2 \,{\rm GeV}^2$.

The readers can check these  guesses  by  explicit 
fitting procedure. To this end, one should  plot the combination 
$f_{\pi}^2F_{\pi}(Q^2)$ as a function of the Borel parameter $M^2$
for different values of the effective threshold $s_0$. 
As a ``true'' value of $s_0$ one can take the value for which 
the curve is constant as $M^2 \to \infty$. One would observe
that the value of $s_0$ obtained in this way slowly grows with
$Q^2$, from $0.6 \, {\rm GeV}^2$ for 
$Q^2=0.5 \,{\rm GeV}^2$ to $1.0 \,{\rm GeV}^2$ 
for $Q^2=3\, {\rm GeV}^2$. The growth of $s_0$ reflects 
the fact that, when the $Q^2$ value
increases, the condensate contributions 
remain constant (with the quark term even slowly growing)  
whereas the perturbative term decreases.
The constancy of the condensate contributions is an artifact
of our approximation. If one includes operators of higher
dimensions in the OPE, there appear terms diminishing the total
condensate contribution for large $Q^2$. These terms, however,
have the structure $(Q^2/M^2)^n$, and one should resum them
to get a reasonable result. 
However, if one restricts the analysis to the region $0.5\, {\rm GeV}^2 <
Q^2 < \, 3 \, {\rm GeV}^2$, the predictions of our sum rule 
agree with existing pion form factor data within 
$10\% - 20\%$ accuracy.

 \subsection{ Local quark-hadron duality for the pion form factor}

The approach described above allows to obtain
 the form factor at different
values of $Q^2$  by independent fitting  procedure, point
by point. But it is tempting  to see the $Q^2$-curve as a whole.
This can be done by looking at the $F_{\pi}(Q^2)$-curves 
at fixed $s_0$ for different values of the Borel parameter $M^2$. 
For $M^2 > 1.5\,{\rm GeV}^2$, all the curves 
are close to each other, approaching the limiting ($s_0$-dependent)
form as $M^2 \to \infty$. The nature of this convergence is very simple:
the condensate terms  disappear from the sum rule 
in the $M^2 \to \infty$ limit, and the limiting curve 
corresponds to the local duality relation:
\be
f_{\pi}^2F_{\pi}(Q^2) = \frac1{\pi^2}\int_0^{s_0}ds_1
\int_0^{s_0}\rho^{pert}(s_1,s_2,Q^2) \, ds_2 .
\end{equation}
Note, that the function $\rho^{pert}(s_1,s_2,Q^2)$ 
describes the transition from a free-quark $(\bar qq$) 
state with invariant mass $s_1$ 
to a similar state with mass $s_2$. The local duality relation
states, essentially, that one can calculate the pion form factor
by averaging the form factors of such transitions over the 
appropriate duality interval $s_0$. In the two-point 
case, the local duality  produces the relation between the pion
decay constant $f_{\pi}^2$ and the duality interval $s_0$
\be
s_0= 4 \pi^2 f_{\pi}^2 \approx 0.7\, {\rm GeV}^2.
\end{equation}
We recall that the duality interval is the effective threshold
for production of higher resonances, and is very natural that
it is just in the middle between $m_{\pi}^2 \approx 0$
and $m_{A_1}^2 \approx 1.6 \, {\rm GeV}^2$. 
Using the explicit form for $\rho^{pert}(s_1,s_2,Q^2)$, we obtain\cite{ffsr} 
\be
F_{\pi}(Q^2)|^{local \,duality} = 1 - \frac{1+6s_0/Q^2}{(1+4s_0/Q^2)^{3/2}}.
\end{equation}
This formula gives a correct value for the form factor at $Q^2=0$, 
but the predicted slope is wrong:
\be
F_{\pi}(Q^2)|^{local \,duality} = 1-\frac3{4}\frac{Q}{\sqrt{s_0}}+\ldots .
\end{equation}
This is just the same situation we encountered 
applying the local duality to the form factor 
of the lowest oscillator state. The reason is that there is 
a serious mismatch between perturbative and physical spectral 
densities: the perturbative density is concentrated in the narrow 
stripe $|s_1 - s_2| < Q$ near the $s_1 = s_2$ line, while 
physical density, in addition to the elastic form factors
(points at the  $s_1 = s_2$ line), contains also transition form 
factors (line $s_1 =0,  s_2>s^{(3\pi\,threshold)}, etc. $) corresponding 
to the regions very far from the $s_1 = s_2$ line. In other words,
our model for the hadronic spectrum is too rough
at small nonzero $Q^2$. The perturbative density in this region
can be approximated by
\be
\rho^{pert}(s_1,s_2,Q^2) \sim \frac1{Q} \, \theta (|s_1 - s_2| < Q).
\end{equation}
Integrating it over any line $s_1+s_2=$const., one obtains  a result
analytic in $Q^2$. In our model, however, there are small triangles
not included into the integration region. Their area is $\sim Q^2$, 
and the density is $\sim 1/Q$, so the resulting contribution 
(to be subtracted
from $1$) is $O(Q)=O(\sqrt{Q^2})$. 
To avoid this, one can try a 
``triangle'' model for the higher states: ``pion + (continuum outside
the triangle formed by the line $(s_1+s_2=S_0)$''. The local duality
relation in this case gives\cite{ffsr} 
\be
F_{\pi}(Q^2)^{triangle\,l.d.} = \frac{S_0}{8\pi^2 f_{\pi}^2 
(1+{Q^2}/{2S_0})^2}.
\end{equation}
To get the correct normalization $F_{\pi}(0)=1$ one should take
$S_0=8\pi^2 f_{\pi}^2=2s_0$. However, the slope in this case is 
too small 
$F_{\pi}^{tr.l.d.}(Q^2)|_{S_0=2s_0}=1-Q^2/2s_0$, by factor 2 smaller than
the experimental one, and the curve goes well above both the experimental
points and the curve corresponding to the ``squared'' form of the 
local duality. To match with the latter, one should take $S_0=\sqrt{2}s_0$. 
In this case the area of the local  duality region 
is the same.  Outside the small-$Q^2$ region, 
there is good agreement between
the two local duality curves and the data approximated by the fit
$F_{\pi}^{fit} (Q^2)\approx 1/(1+Q^2/0.47)$.

\subsection{ One-gluon exchange contribution} 

The contribution
into the  pion
form factor we calculated using the lowest 
order $O(\alpha_s^0)$ term
in the perturbative spectral density corresponds to the so-called
``soft diagram''. Since this contribution 
decreases as $1/Q^4$ for large $Q^2$, it is 
normally ignored in the perturbative QCD analysis concentrated on the study
of the ``hard gluon exchange diagram'' that has the $1/Q^2$ asymptotic
behavior dictated by the quark counting rules. 

However,
 the soft contribution as given, e.g.,
 by the local quark-hadron duality, is very close to the experimental data 
leaving not much place for any other contribution.
To estimate the contribution due to the ``hard gluon'' exchange
diagrams, one should  include the $O(\alpha_s(Q^2)$-terms in the 
perturbative spectral density. To simplify the analysis, one can use
the local duality approximation. Still, the two-loop
calculation is rather complicated, but the results, with 
 a rather high accuracy, can be approximated by a simple 
interpolation
\begin{equation}
\frac1{\pi^2f_\pi^2}\int_0^{s_0}ds_1\int_0^{s_0}
ds_2\Delta\rho^{pert}(s_1,s_2,q^2) \approx \frac{\alpha_s}{\pi} 
 \frac{1}{1+Q^2/2s_0} 
  \label{eq:hardgluonld}
\end{equation}
between the $Q^2=0$ value (related by the Ward identity to the $ O(\alpha_s)$
 term of the 2-point correlator) and the asymptotic behavior
$F_{\pi}^{as}(Q^2)=8\pi \alpha_s (Q^2) f_{\pi}^2/Q^2$.

\subsection{Pion form factor at small $Q^2$} 

As we discussed earlier, if the QCD sum rule derived
in the moderate-$Q^2$ region
is   simple-mindedly extrapolated   to the point $Q^2=0$, 
one obtains the expression 
\begin{equation}
f_{\pi}^2F_{\pi}(0)\, \stackrel{?}{=} \,\frac{m^2}{4\pi^2} 
\left(1-e^{-s_0/m^2}\right) 
+\frac{\alpha_s\langle GG \rangle}{24\pi m^2} +
\frac{52}{81}\frac{\pi\alpha_s \langle \bar qq\rangle^2}{m^4}
\end{equation}
that differs from the sum rule for $f_{\pi}^2$ 
\be
f_{\pi}^2 = \frac{m^2}{4\pi^2} 
\left(1-e^{-s_0/m^2}\right) 
+\frac{\alpha_s\langle GG \rangle}{12\pi m^2} +
\frac{176}{81}\frac{\pi\alpha_s \langle \bar qq\rangle^2}{m^4}.
\end{equation}
The condensate terms have the coefficients essentially smaller
than it is necessary to enjoy the simple Ward identity
prediction $F_{\pi}(Q^2)=1$. The latter requires that the QCD sum rule
for $f_{\pi}^2F_{\pi}(0) $ should coincide with that for 
$f_{\pi}^2$. The Ward identity, being just the statement that
the total electric charge of the pion is equal to that
of the electron, should not be violated. Hence there should
exist the additional condensate contributions ``visible'' 
at $Q^2=0$ (to satisfy the Ward identity) and dying rapidly
for $Q^2 > m_{\rho}^2$ (to reproduce the intermediate-$Q^2$
sum rule). 
 
  \begin{figure}[t]
\mbox{
   \epsfxsize=8cm
 \epsfysize=3cm
\hspace{2cm} 
  \epsffile{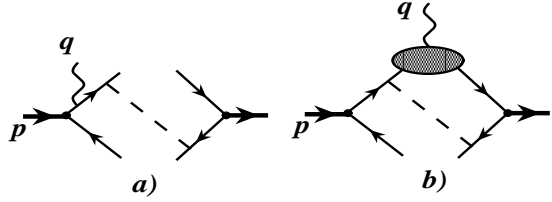}  }
{\caption{\label{fig:biloc} 
Gauge-invariant set of $\langle \bar qq \rangle^2$ 
diagrams.}}
\end{figure}

To get a feeling about how these terms might look like, consider
the $\alpha_s\langle \bar qq \rangle^2$ diagrams.
To enjoy the consequences of   the Ward identity, one should have a 
gauge invariant set of diagrams. This can be achieved by inserting
the photon vertex into all possible lines, including also the 
``external'' lines corresponding to quarks going into  vacuum.
However, in perturbation theory, this brings in terms $\sim 1/Q^2$
singular in the $Q^2 \to 0$ limit. Such a ``coefficient function''  
cannot be calculated perturbatively. Thus, one should separate the
large- and small-momentum parts of the diagram 
\be
T(p^2,q^2)= C(p^2) \otimes \Pi(q^2)
\end{equation}
and treat the 
$Q^2$-dependent part as a correlator of the original electromagnetic
current $J$ and some local operator constructed from the quark
fields going ``into vacuum'' (see Fig.\ref{fig:biloc}b):
\be
\Pi(q^2) \sim \int \, e^{iqx} \langle T(J_{\mu}(0) {\cal O}(x)) \rangle d^4x.
\end{equation}
The question now is whether it is possible to calculate contributions 
like the correlators $\Pi(q^2)$
at small $q$?  There are essentially two types of these correlators depending
on the type of the operator ${\cal O}$.
 First, one should take into account that under 
 the $T$-product sign, one cannot throw out the
operators apparently vanishing due to the equations of motion, 
{\em e.g.,} those containing $\slash \hspace{-2.5mm}D$ acting on the quark field, 
since there appear contact terms. In this case $\Pi(q^2)$ is a
constant proportional to the quark condensate $\langle \bar qq \rangle$.
 In all other cases one can (approximately) calculate $\Pi(q^2)$
using the following QCD sum rule strategy:
 \\ $\bullet$  Calculate $\Pi(q^2)$ for large $Q^2$ using the operator
product expansion. This normally gives
$$\Pi(q^2=-Q^2) \sim \frac{\langle \bar qq \rangle}{Q^2}+ 
k \frac{\langle \bar qD^2q \rangle}{Q^4}+ \ldots . $$
\\ $\bullet$  Use the dispersion relation 
$$\Pi(Q^2)= \frac1{\pi} \int_0^{\infty} \frac{\rho(s) \,ds}{s+Q^2}.$$
\\ $\bullet$  Assume an appropriate ansatz for the spectral density:
$$ \rho(s) = \pi A \delta(s-m_{\rho}^2) + \pi B \delta (s-m_R^2).$$
Since the perturbative density is zero in this case, the higher
states (lying above the $\rho$-meson) are approximated by
an effective resonance $R$.
\\ $\bullet$  Requiring the best agreement between the two representations
for $\Pi(Q^2)$, extract the values of the constants $A,B$ and $m_R^2$.
\\ $\bullet$  The result is 
$$ \Pi(Q^2) = \frac{A}{Q^2+m_{\rho}^2} + \frac{B}{Q^2+m_R^2},$$
and one can use it for small $Q^2$.

In this way one can obtain the QCD sum rule for the pion 
form factor valid in the low-$Q^2$ region\cite{smallq2}:
\ba
f_{\pi}^2F_{\pi}(Q^2) = ``perturbative\, part" 
+\frac{\alpha_s\langle GG \rangle}{12\pi m^2}\left[ \frac1{2} +
\frac{m_{\rho}^2}{2(m_{\rho}^2+Q^2)}\right]+ \nonumber \\ +
\frac{16}{81}\frac{\pi\alpha_s \langle \bar qq\rangle^2}{m^4}
\left[4+6\left(\frac{1.6\, m_{\rho}^2}{m_{\rho}^2+Q^2}-
\frac{0.6\, m_R^2}{m_R^2+Q^2}\right)+ \frac{m_{\rho}^2}{m_{\rho}^2+Q^2}
+\frac{Q^2}{m^2}\right].
\ea
 This sum rule possesses all the required properties:
 \\ $\bullet$  For $Q^2=0$ its left hand side exactly coincides with
that of the $f_{\pi}^2$ sum rule. 
\\ $\bullet$  additional terms die out when $Q^2$ gets large.
\\ $\bullet$  The curve for the pion form factor calculated
with the help of this sum rule in the region of small $Q^2$,
at the matching point $Q^2=m_{\rho}^2$, agrees within 10\% 
with the curve obtained from the moderate-$Q^2$ sum rule.
\\ $\bullet$  the pion charge radius, extracted from the low-$Q^2$
sum rule, 
$$\langle r_{\pi}^2\rangle^{1/2}|_{SR} = 0.66 \pm 0.03 \,fm$$
is in good agreement with the experimental value 
$$ \langle r_{\pi}^2\rangle^{1/2}|_{exp.} = 0.636 \pm 0.036 \,fm \,.$$

\section{Conclusions}

Thus, the QCD sum rules give an accurate description of the pion form
factor in the whole region where the  data exist. 
They allow one to trace its behavior  from the normalization point $Q^2=0$,
through the low-$Q^2$ region, to the region of moderately large $Q^2$.
Everywhere the QCD sum rule results are in 
agreement with experiment within the accuracy of the method.
The QCD sum rules unambigously demonstrate that, in the 
experimentally accessible region, the form factor is dominated
by the soft  contribution, which is not calculable  within a purely perturbative
approach. They show that the one-gluon-exchange mechanism, 
though dominant in the asymptotic limit, is of minor importance
in the region $Q^2<4 \,{\rm GeV}^2$ and, maybe, till even larger values of $Q^2$.

\section*{Acknowledgements}

This work was supported
by the US Department of Energy 
under contract DE-AC05-84ER40150.

\end{document}